\documentclass[11pt]{article}
\pdfoutput=1

\usepackage{appendix}

\usepackage{amssymb,amsmath,bm}
\usepackage{cite}
\usepackage{graphicx}
\usepackage{color}

\def\slash#1{\setbox0=\hbox{$#1$}\dimen0=\wd0
      \setbox1=\hbox{/} \dimen1=\wd1 \ifdim\dimen0>\dimen1
      \rlap{\hbox to \dimen0{\hfil/\hfil}} #1                        \else
      \rlap{\hbox to \dimen1{\hfil$#1$\hfil}}
      /   \fi}

\renewcommand{\theequation}{\thesection.\arabic{equation}}

\newcommand{\lsim}{
\mathrel{\hbox{\rlap{\hbox{\lower4pt\hbox{$\sim$}}}\hbox{$<$}}}}

\newcommand{\gsim}{
\mathrel{\hbox{\rlap{\hbox{\lower4pt\hbox{$\sim$}}}\hbox{$>$}}}}

\allowdisplaybreaks[2]

\long\def\symbolfootnote[#1]#2{\begingroup%
\def\thefootnote{\fnsymbol{footnote}}\footnote[#1]{#2}\endgroup} 

\begin{document}
\begin{titlepage}
\vspace*{-1.0truecm}

\begin{flushright}
IPPP/11/70 \\
DCPT/11/140
\\
(incl.\ erratum)
\end{flushright}

\vspace{7em}

\begin{center}
\boldmath
{\Large\textbf{Form Factors for  $\Lambda_b \to \Lambda$ Transitions in SCET}}
\unboldmath
\end{center}

\vspace{3em}

\begin{center}
{\bf Thorsten Feldmann}\symbolfootnote[3]{Address after 1~Nov~2011: {\sl Theoretische Elementarteilchenphysik, 
Naturwissenschaftlich Techn.\ Fakult\"at,
Universit\"at Siegen, 57068 Siegen, Germany}, email:feldmann@hep.physik.uni-siegen.de},
{\bf Matthew W\,Y Yip}\symbolfootnote[4]{email:m.w.yip@durham.ac.uk}
\vspace{0.4truecm}

{\footnotesize
{\sl IPPP, Department of Physics, University of Durham, Durham DH1 3LE, UK}\vspace{0.2truecm}
}

\end{center}

\vspace{5em}
\begin{abstract}
\noindent

We present a systematic discussion of $\Lambda_b \to \Lambda$ transition form factors
in the framework of soft-collinear effective theory (SCET). The universal soft form
factor, which enters the symmetry relations in the limit of large recoil energy,
is calculated from a sum-rule analysis of a suitable SCET correlation function. 
The same method is applied to derive the leading corrections from hard-collinear
gluon exchange at first order in the strong coupling constant. We present numerical 
estimates for form factors and form-factor ratios and their impact on decay observables in
$\Lambda_b \to \Lambda \mu^+\mu^-$ decays.

\end{abstract}

\clearpage

\thispagestyle{empty}

\tableofcontents

\end{titlepage}
\setcounter{page}{0}
\pagenumbering{arabic}


\section{Motivation}

The decays $\Lambda_b \to \Lambda \ell^+\ell^-$  offer the possibility to study 
rare semi-leptonic and radiative $b \to s$ transitions within the Standard Model (SM) and beyond. The observables in the baryonic
transitions provide complementary phenomenological information compared to the corresponding 
mesonic or inclusive decays, see e.g.\ \cite{Chen:2001zc,Liu:2011em,Aliev:2010uy,Wang:2009hra,Hiller:2001zj,Oliver:2010im,Colangelo:2007jy,Hiller:2007ur,Mott:2011cx}.
The decay  $\Lambda_b \to \Lambda \mu^+\mu^-$ has been recently measured by the CDF collaboration \cite{Aaltonen:2011qs}
with a branching ratio of the order $10^{-6}$.

Theoretical predictions for the exclusive decay matrix elements require non-perturbative hadronic input.
To first approximation, this can be parametrized in terms of baryonic transition form factors for vector,
axial-vector and tensor currents. The number of independent form factors drastically reduces in the 
limit of infinitely heavy $b$-quark mass, exploiting the approximate symmetries in heavy-quark effective theory (HQET),
see e.g.\ \cite{Mannel:1990vg,Mannel:1997xy}. 
Additional simplification is expected in
the kinematic limit of large recoil energy, $E_\Lambda \to \infty$, where the number of
independent form factors is known to be reduced further \cite{Charles:1998dr}, and part of the corrections 
to this limit should factorize in terms of process-independent hadronic quantities (light-cone distribution amplitudes, 
LCDAs,
\cite{Liu:2008yg,Ball:2008fw}) and perturbative interaction kernels, in a similar way as it has been discussed
for the analogous mesonic transitions \cite{Beneke:2000wa}.
The non-perturbative calculation of the remaining hadronic transition form factors can, for instance,
be obtained from QCD sum rules. In the limit of large recoil energy, a systematic expansion
in the heavy-quark mass is achieved in the framework of soft-collinear effective theory 
(SCET \cite{Bauer:2000yr,Beneke:2002ph}), where one studies the spectrum of correlation functions between
the decay current and an interpolating current with the quantum numbers of the light hadron \cite{DeFazio:2005dx}
(see also \cite{Khodjamirian:2006st}).

The aim of this paper is to provide a systematic analysis of $\Lambda_b \to \Lambda$ form factors, starting
from the symmetry relations in the heavy-quark/large-energy limit. To this end, in the next section, we
will present a convenient definition of the 10 independent physical form factors, in terms of which the
HQET and SCET symmetry relations look particularly simple. In the following section~\ref{sec:SR},
we derive the leading expressions for the universal (``soft'') form factor $\xi_\Lambda$ from a sum-rule analysis
of an appropriate correlation function in SCET, involving the LCDAs of the $\Lambda_b$ baryon. 
The same method is used to calculate the leading correction $\Delta \xi_\Lambda$ to the form-factor symmetry
relations that arise from hard-collinear gluon exchange. In contrast to the mesonic case,
one of the light spectator quarks still does not take part in the hard-scattering process, and
therefore the corresponding effect could not be calculated in the framework of QCD factorization.
(A similar discussion has been led for the electromagnetic form factors of the nucleon in
\cite{Kivel:2010ns}.) The sum-rule expressions are analysed numerically in section~\ref{sec:numerics}.
We focus on the theoretical uncertainty related to various hadronic input parameters entering
the estimate for the soft form factor $\xi_\Lambda$. Part of these uncertainties drops out in the
ratio $\Delta \xi_\Lambda/\xi_\Lambda$. We also provide estimates for the partial branching fractions 
(transverse and longitudinal rate, forward-backward asymmetry)
of $\Lambda_b \to \Lambda \mu^+\mu^-$ in the large recoil (small $q^2$) limit, before
we conclude. Finally, in our appendix, we collect the expressions for the double-differential
$\Lambda_b \to \Lambda \mu^+\mu^-$ decay rates, and discuss an alternative form-factor basis
that is optimized for a systematic discussion of power corrections to the symmetry relations.
We also extract the hard vertex corrections to the $\Lambda_b \to \Lambda$ form factors 
arising from the matching of the decay currents from QCD onto SCET, and we identify 5 
form-factor relations that are unaffected by short-distance ${\cal O}(\alpha_s)$ corrections. 
Finally, we summarize the relevant information on baryon LCDAs and briefly comment on
a simplified set-up with elementary light di-quark fields in the light and heavy baryon.

\section{$\Lambda_b \to \Lambda$ Form Factors} \label{sec:ff}

In the following, we provide some useful definitions for $\Lambda_b \to \Lambda$ form factors
that aim to improve previous definitions, as discussed for instance in \cite{Chen:2001zc,Wang:2009hra},
in two aspects: (i) the form factors are defined from a helicity basis, (ii) the form factors
are normalized to the limit of point-like hadrons. As a result, our form factor convention leads to
rather simple expressions for partial rates, unitarity bounds (cf.\ \cite{Boyd:1997qw,Bharucha:2010im}) and symmetry relations 
in the HQET or SCET limit.

\subsection{Helicity-Based Form-Factor Parametrization}

The form factors for $\Lambda_b \to \Lambda$ transitions can be parametrized as follows. 
Starting with the vector and scalar decay currents, we have ($q=s(x)$ and $b=b(x)$ denote the light
and heavy quark fields in the $b \to s$ transitions) 
\begin{align}
 \langle \Lambda(p',s')| \bar q \, \gamma_{\mu} \, b|\Lambda_b(p,s)\rangle &= \bar u_{\Lambda}(p',s') 
 \left\{ 
  f_0(q^2) \, (M_{\Lambda_b}-m_\Lambda) \, \frac{q_\mu}{q^2}  
\right.
\cr 
& \qquad  \left.
 + f_+(q^2) \, \frac{M_{\Lambda_b}+m_{\Lambda}}{s_+} \,
 \left(p_\mu + p_\mu' - \frac{q_\mu}{q^2} \, (M_{\Lambda_b}^2-m_{\Lambda}^2) \right)
\right.
\cr 
&\qquad \left.
+ f_\perp(q^2) \left( \gamma_\mu - \frac{2 m_\Lambda}{s_+} \, p_\mu - \frac{2  M_{\Lambda_b}}{s_+} \, p '_\mu
  \right)
 \right\} u_{\Lambda_b}(p,s) \,,
\end{align}
where we have defined
\begin{align}
 s_\pm &= (M_{\Lambda_b}\pm m_\Lambda)^2-q^2 \,.
\end{align}
At vanishing momentum transfer, $q^2 \to 0$,
one further has the kinematic constraint 
\begin{align}
 f_0(0) &= f_+(0)  \,.
\end{align}
The individual form factors are defined in such a way that 
they correspond to time-like (scalar), longitudinal and transverse polarization with respect to the 
momentum-transfer $q^\mu$ for $f_0$, $f_+$ and $f_\perp$, respectively (cf.\ \cite{Boyd:1997qw,Bharucha:2010im}).
The normalization is chosen in such a way that for
$$
  f_0, f_+, f_\perp \to 1 \,, 
$$
one recovers the expression for a transition between point-like baryons, i.e.\
$\langle \Lambda| \bar q \,\Gamma \,  b|\Lambda_b\rangle \to \bar u_\Lambda \, \Gamma \, u_{\Lambda_b}$.
The form factor $f_0$ is also obtained from the scalar decay current via the equations of motion (e.o.m.),
\begin{align}
 \langle \Lambda(p',s')| \bar q \,  b|\Lambda_b(p,s)\rangle &= \frac{q^\mu}{M_b-m_q}  \,\langle \Lambda(p',s')| \bar q \, \gamma_{\mu} \, b|\Lambda_b(p,s)\rangle
\cr & =
 f_0(q^2) \, \frac{M_{\Lambda_b}-m_\Lambda}{M_b-m_q} \, \bar u_{\Lambda}(p',s') \,  u_{\Lambda_b}(p,s) \,.
\end{align}
The expression for the  axial-vector and pseudo-scalar currents can be obtained
by appropriately changing the relative sign between the light and heavy baryon mass, and we thus define
\begin{align}
 \langle \Lambda(p',s')| \bar q \, \gamma_\mu \gamma_5 \, b|\Lambda_b(p,s)\rangle &= {} - \bar u_{\Lambda}(p',s') \gamma_5
 \left\{ 
  g_0(q^2) \, (M_{\Lambda_b}+m_\Lambda) \, \frac{q_\mu}{q^2}  
\right.
\cr 
& \qquad  \left.
 + g_+(q^2) \, \frac{M_{\Lambda_b}-m_{\Lambda}}{s_-}
\left( p_\mu + p_\mu' - \frac{q_\mu}{q^2} \, (M_{\Lambda_b}^2-m_{\Lambda}^2) \right)
\right.
\cr 
&\qquad \left.
+ g_\perp(q^2) \left( \gamma_\mu + \frac{2 m_\Lambda}{s_-} \, p_\mu - \frac{2 M_{\Lambda_b}}{s_-} \,  p'_\mu
  \right)
 \right\} u_{\Lambda_b}(p,s) \,,
\cr &
\end{align}
with the kinematic constraint $g_0(0) = g_+(0)$ at $q^2 \to 0$,
and
\begin{align}
 \langle \Lambda(p',s')| \bar q \, \gamma_5 \,  b|\Lambda_b(p,s)\rangle &= \frac{q^\mu}{M_b+m_q}  \,\langle \Lambda(p',s')| \bar q \, \gamma_5 \gamma_{\mu} \, b|\Lambda_b(p,s)\rangle
\cr & =
 g_0(q^2) \, \frac{M_{\Lambda_b}+m_\Lambda}{M_b+m_q} \,\bar u_{\Lambda}(p',s') \, \gamma_5 u_{\Lambda_b}(p,s) \,.
\end{align}
Finally, for the tensor and pseudo-tensor current, we write
\begin{align}
&  \langle \Lambda(p',s')| \bar q \, i\sigma_{\mu\nu} q^\nu \, b|\Lambda_b(p,s)\rangle 
\cr 
&= {} -  \bar u_{\Lambda}(p',s') 
 \left\{  
 h_+(q^2) \, \frac{q^2}{s_+}
\left( p_\mu + p_\mu' - \frac{q_\mu}{q^2} \, (M_{\Lambda_b}^2-m_{\Lambda}^2) \right)
\right.
\cr 
&\qquad \left.
+(M_{\Lambda_b}+m_\Lambda) \, h_\perp(q^2) 
\left( \gamma_\mu -  \frac{2  m_\Lambda}{s_+} \, p_\mu - \frac{2M_{\Lambda_b}}{s_+} \, p'_\mu 
  \right)
 \right\} u_{\Lambda_b}(p,s) \,,
\end{align}
and
\begin{align}
& \langle \Lambda(p',s')| \bar q \, i\sigma_{\mu\nu} \gamma_5 q^\nu \, b|\Lambda_b(p,s)\rangle
\cr 
&= {} -  \bar u_{\Lambda}(p',s') 
 \gamma_5 \left\{  
 \tilde h_+(q^2) \, \frac{q^2}{s_-}
\left( p_\mu + p_\mu' - \frac{q_\mu}{q^2} \, (M_{\Lambda_b}^2-m_{\Lambda}^2) \right) 
\right.
\cr 
&\qquad \left.
+ (M_{\Lambda_b}-m_\Lambda) \,\tilde h_\perp(q^2) \left( \gamma_\mu + 
 \frac{2 m_\Lambda}{s_-} \, p_\mu - \frac{2 M_{\Lambda_b}}{s_-} \, p'_\mu
  \right)
 \right\}  u_{\Lambda_b}(p,s) \,.
\end{align}
Again, the normalization of the form factors $h_{\perp,+},\tilde h_{\perp,+}$
has been fixed by the case of point-like hadrons. 
This makes 10 independent form factors for the general case, after the e.o.m.\
have been taken into account.\footnote{For convenience, we summarize in Appendix~\ref{app:otherff} the relations
of the 10 helicity form factors to the various form factors defined in \cite{Chen:2001zc}.}
In terms of the helicity form factors, the differential decay width for $\Lambda_b \to \Lambda \mu^+\mu^-$ takes a particularly simple form, 
see Appendix~\ref{app:gammas}. An alternative parametrization, which is based on the large and small
projections of energetic or massive fermion spinors, can be found in Appendix~\ref{app:altpar}.

\subsection{HQET Limit}

The number of independent $\Lambda_b \to \Lambda$ form factors reduces considerably in the
heavy quark limit, $M_b \to \infty$ (see e.g.\ \cite{Mannel:1997xy}), when we use the
heavy-baryon velocity $v^\mu$ to project onto the large spinor components $h_v^{(b)}=\slash v \, h_v^{(b)}$ 
of the heavy $b$-quark field,
\begin{align}
 \langle \Lambda(p',s')| \bar q \, \Gamma \, b | \Lambda_b(p,s) \rangle 
&\to  \langle \Lambda(p',s')| \bar q \, \Gamma \, h_v^{(b)} | \Lambda_b(v,s) \rangle
\cr
& \simeq \bar u_\Lambda(p',s') \left( A(v \cdot p')+ \slash v \, B(v \cdot p') \right) \Gamma \, u_{\Lambda_b}(v,s) \,. 
\label{hqlim}
\end{align}
Here $\Gamma$ is an arbitrary Dirac matrix, and $p^\mu=M_{\Lambda_b} v^\mu \simeq M_b v^\mu$. Furthermore,
$|\Lambda_b(v,s)\rangle$ is a heavy-baryon state, and $u_{\Lambda_b}(v,s)=\slash v \, u_{\Lambda_b}(v,s)$ a 
heavy-baryon spinor in HQET.
In the heavy-quark limit, $m_\Lambda, v\cdot p' \ll M_b$, the helicity form factors
are related to the two HQET form factors in (\ref{hqlim}) as follows,
\begin{align}
\mbox{\underline{small recoil:}} \quad \  f_0(q^2) 
\simeq g_+(q^2) 
\simeq g_\perp(q^2)
\simeq \tilde h_+(q^2)
\simeq \tilde h_\perp(q^2) 
\simeq
& \
 A (v\cdot p') + B (v\cdot p') \,,
\cr 
 g_0(q^2) 
\simeq f_+(q^2) 
\simeq f_\perp(q^2)
\simeq h_+(q^2)
\simeq h_\perp(q^2)  
\simeq & \ A (v\cdot p') - B (v\cdot p') \,,
\label{hqetrelation}
\end{align}
with $q^2=M_{\Lambda_b}^2 - 2 M_{\Lambda_b} \, v \cdot p' + m_\Lambda^2$.

\subsection{SCET Limit and Hard-Scattering Corrections}

In the kinematic region of large recoil energy for the $\Lambda$ baryon in the rest frame of the decaying $\Lambda_b$,
further simplifications arise \cite{Charles:1998dr,Beneke:2000wa}. A formal derivation can
be obtained from soft-collinear effective theory (SCET) \cite{Bauer:2000yr,Beneke:2002ph}.
To this end, we  consider
the matrix element of the leading current involving the collinear quark field $\xi \equiv \frac{\slash n_-\slash n_+}{4} \, q$
with two light-like vectors $n_-^2 =n_+^2=0$, satisfying $(n_++n_-)/2 = v$ and $n_- \cdot n_+ =2$.
In the large-energy limit, we can further set $p'{}^\mu \simeq (n_+p') n_-^\mu/2$ and $m_\Lambda \to 0$. 
This amounts to
\begin{align}
&  \langle \Lambda(p',s')| \bar \xi W \, \Gamma \, Y^\dagger h_v^{(b)} | \Lambda_b(v,s) \rangle
 = \bar u_\Lambda(p',s')  \left( A(q^2)+ \slash v \, B(q^2) \right) \frac{\slash n_+\slash n_-}{4} \, \Gamma \,  u_{\Lambda_b}(v,s)
\cr 
& \qquad = A(q^2) \,  \bar u_\Lambda(p',s') \, \frac{\slash n_+\slash n_-}{4} \, \Gamma \,  u_{\Lambda_b}(v,s)
+ B(q^2)  \,  \bar u_\Lambda(p',s') \, \frac{\slash n_-}{2} \, \Gamma \,  u_{\Lambda_b}(v,s) \,.
\end{align}
Here $W$ ($Y$) are  Wilson lines in SCET that render the definition of the form factors
invariant under collinear (soft) gauge transformations. In the following, we will always drop
the Wilson lines (which corresponds to light-cone gauges for collinear and soft gluon fields).
Exploiting the (approximate) equations of motion for $\bar u_{\Lambda}(p',s') \, \slash n_- \simeq 0$,
 this simplifies to
\begin{align}
&  \langle \Lambda(p',s')| \bar \xi \, \Gamma \, h_v^{(b)} | \Lambda_b(v,s) \rangle
 \simeq  \xi_\Lambda(n_+p') \, \bar u_\Lambda(p',s') \, \Gamma \,  u_{\Lambda_b}(v,s) \,,
\label{soft}
\end{align}
where $\xi_\Lambda(n_+p')$ corresponds to $A(v\cdot p')$ in (\ref{hqetrelation}) and 
defines the so-called ``soft'' $\Lambda_b \to \Lambda$ form factor,
while the contribution from $B(v\cdot p')$ is negligible.
In the SCET limit, $n_+p' \sim M_{\Lambda_b}$, all helicity form factors are thus equal to $\xi_\Lambda(n_+p')$, 
\begin{align}
\mbox{\underline{large recoil:}} \qquad \ & f_0(q^2) 
 \approx f_+(q^2)  \approx f_\perp(q^2) \approx h_+(q^2) \approx h_\perp(q^2)
\cr 
 \approx \ & g_0(q^2) \approx g_+(q^2)  
\approx g_\perp(q^2)  \approx  
\tilde h_+(q^2) \approx \tilde h_\perp(q^2) \ 
            \approx \ \xi_\Lambda(n_+p') \,,
\end{align}
with $q^2 = M_{\Lambda_b}^2- M_{\Lambda_b} \, n_+p' + m_{\Lambda}^2 \left(1- \frac{M_{\Lambda_b}}{n_+p'} \right)$.


The leading corrections to the form factor relations from hard-collinear gluon exchange can be
described by a form factor term that takes into account the corresponding sub-leading currents
in SCET, which contain one additional (transverse) hard-collinear gluon field \cite{Beneke:2002ph}. 
If we neglect additional hard vertex corrections for simplicity, the form factors relate
to matrix elements of local SCET currents. In  the limit $M_b \to \infty$, $(n_+p')\to \infty$,
these matrix element can again be described by a single form factor $\Delta \xi_\Lambda$, which we define
by 
\begin{align}
  \langle \Lambda(p',s')| \bar \xi \, \tilde \Gamma \, g A_\mu^\perp  \, h_v^{(b)} | \Lambda_b(v,s) \rangle
 &\equiv  M_{\Lambda_b} \, \Delta \xi_\Lambda(n_+p') \ \bar u_\Lambda(p',s') \, \gamma_\mu^\perp \,\tilde \Gamma \,  u_{\Lambda_b}(v,s) \,,
\label{deltasoft}
\end{align}
where the basis of independent Dirac matrices can be reduced to $\tilde \Gamma = \frac{\slash n_+}{2} \, \{1,\gamma_\nu^\perp,\gamma_5 \}$.
Here we have exploited again that, due to the heavy-quark spin symmetry, the Dirac matrix in the effective-theory 
decay current couples trivially to the heavy baryon spinor.
The matching of the various decay currents in QCD onto the SCET currents is process-independent and 
can be taken into account by appropriate Wilson coefficients.
For convenience, we have summarized the relevant results in Appendix~\ref{app:corr}.


\section{Sum rules in SCET}

\label{sec:SR}

\begin{figure}[tpb]
 \begin{center}
  \includegraphics[height=0.18\textheight]{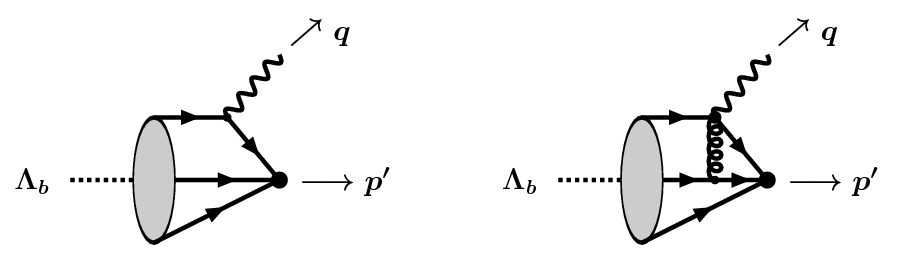}
 \end{center}
\caption{\small \small Leading diagrams for SCET correlation functions involving 
 the soft form factor $\xi_\Lambda$ (left) and the form factor $\Delta \xi_\Lambda$
 for the hard-scattering corrections in the large-recoil limit.}\label{fig} 
\end{figure}

Our next aim is to obtain non-perturbative estimates for the form factors 
$\xi_\Lambda$ and $\Delta \xi_\Lambda$
in the large-recoil limit, following the analogous calculation 
as for the $B \to \pi(\rho)$ form factors from SCET sum rules in \cite{DeFazio:2005dx}.
The leading diagrams for the calculation of the respective correlation functions
are shown in Fig.~\ref{fig}.

\subsection{Soft form factor}

\label{sec:soft}

We start with a correlation function, where the $\Lambda$ baryon in the final state
is replaced by an interpolating current sharing the same quantum numbers. 
We choose 
\begin{align} 
 J_\Lambda(x) &\equiv \epsilon^{abc} \left(u^a(x) \, C \,\gamma_5 \slash n_+  \, d^b(x)\right)  s^c(x) \,,
\end{align}
which is normalized by the matrix element 
\begin{align}
 \langle 0 | \frac{\slash n_\mp\slash n_\pm}{4} \, J_\Lambda(0) |\Lambda(p',s')\rangle &= 
(n_+p') \, f_\Lambda \,\frac{\slash n_\mp\slash n_\pm}{4} \, u_\Lambda(p',s') \,,
\end{align}
and thus corresponds to a leading term in the large-energy limit.
A sum-rule estimate \cite{Liu:2008yg} gives $f_\Lambda \simeq (6.0 \pm 0.3) \times 10^{-3}~{\rm GeV}^2$ for the involved
decay constant of the $\Lambda$ baryon
(for comparison, for the nucleon  $f_N\simeq 5.6 \times 10^{-3}~{\rm GeV}^2$ has been estimated in \cite{Braun:2000kw}).
The various light-quark fields can be decomposed into soft and hard-collinear fields to match the above current onto SCET.
At tree-level, it is sufficient to calculate the correlation function in QCD and perform the appropriate kinematic limits
for the propagators. 

We now define the correlation function between a weak decay current and
the interpolating current $J_\Lambda$, and consider it as a function of the small (Euclidean) momentum component $(n_-p')<0$
for $p_\perp'\equiv 0$ and fixed large momentum component $(n_+p')$. In order to extract the universal soft form factor $\xi_\Lambda$, 
we consider the projection of a decay current on
the large spinor components for the light and heavy quark fields. We therefore have
\begin{align}
 \Pi_\Lambda(n_-p') & \equiv i \int d^4x \, e^{ip'x} \langle 0| T \left[ \frac{\slash n_- \slash n_+}{4} \, J_\Lambda(x) \left[ \bar s(0) \, \frac{\slash n_+\slash n_-}{4} 
  \, \Gamma \, \frac{1+\slash v}{2} \, b(0) \right] \right] |\Lambda_b(p)\rangle \,.
\end{align}
The time-ordered product of the two currents can be calculated in perturbation theory. The leading diagram just
corresponds to the one shown on the left-hand side in Fig.~\ref{fig}, which refers to the situation where the two
strange-quark fields are contracted to a propagator, while the up- and down-quark merely act as spectators. 
Employing the kinematic limits in the QCD diagram, and performing a Fourier transform such that $\omega_{1,2} = (n_-k_{1,2})$
correspond to the light-cone momenta of the up- and down-quark, the correlation function at leading order is given by 
\begin{align}
 \Pi_\Lambda(n_-p') & \simeq 
  \int  \frac{d\omega_1 \, d\omega_2}{\omega_1+\omega_2 - n_-p' -i\epsilon} \, 
 \langle 0| \epsilon^{abc} \left(u^a(\omega_1) \, C \,\gamma_5 \slash n_+  \, d^b(\omega_2)\right) 
\frac{\slash n_-}{2} \, \Gamma \, b^c_v  |\Lambda_b(v,s)\rangle 
\cr 
 &= f_{\Lambda_b}^{(2)} \int \frac{d\omega_1 \, d\omega_2\, \psi_4(\omega_1,\omega_2) }{\omega_1+\omega_2 - n_-p' -i\epsilon} \
 \frac{\slash n_-}{2} \, \Gamma \, u_{\Lambda_b}(v,s)
\cr 
 &= f_{\Lambda_b}^{(2)} \int_0^\infty \frac{d\omega \, \omega \, \int_0^1 du \, \tilde\psi_4(\omega,u) }{\omega-n_-p' -i\epsilon} \
 \frac{\slash n_-}{2} \, \Gamma \, u_{\Lambda_b}(v,s) \,.
\end{align} 
To arrive at the second and third line, we have used the momentum-space projector for the heavy $\Lambda_b$ baryon, following
from the definition of its light-cone distribution amplitudes as derived in Appendix~\ref{app:DAs}.
To leading order, the result for the correlation function thus only involves the \emph{sum} of the spectator-quark momenta and therefore only 
requires the partially integrated LCDA 
$$
 \phi_4(\omega) \equiv \omega \int_0^1 du \, \tilde\psi_4(\omega,u) \,.
$$

The remaining analysis is then very similar to the $B \to \pi, \rho$ case discussed in \cite{DeFazio:2005dx}.
For the hadronic side of the sum rule, the contribution of the $\Lambda$ baryon to the correlator
is given by
\begin{align}
 \Pi_\Lambda \big|_{\rm res.} & \simeq \sum_{s'} 
  \frac{\slash n_- \slash n_+}{4} \, \frac{ 
\langle 0| J_\Lambda| \Lambda(p',s')\rangle
\langle \Lambda(p',s')| \bar q \, \frac{\slash n_+\slash n_-}{4} 
  \,\Gamma \,  h_v^{(b)} | \Lambda_b(v) \rangle }{m_\Lambda^2- (p')^2}
\cr 
 &= \frac{(n_+p') \, f_\Lambda \, \xi_\Lambda(n_+p')}{m_\Lambda^2- (n_+p')(n_-p')}
 \, \sum_{s'} \frac{\slash n_-\slash n_+}{4} \, u_\Lambda(p',s') \,
 \bar u_\Lambda(p',s') \, \frac{\slash n_+\slash n_-}{4} \, \Gamma\, u_{\Lambda_b}(v,s) 
\cr 
&= \frac{(n_+p') \, f_\Lambda \, \xi_\Lambda(n_+p')}{m_\Lambda^2/(n_+p')- (n_-p')} \, \frac{\slash n_-}{2} \, \Gamma \, u_{\Lambda_b}(v,s)  \,.
\end{align}
Comparing the perturbative and hadronic parts of the sum rule,  subtracting the continuum
(which is modelled by the perturbative result above a threshold parameter $\omega_s$), and
performing a Borel transformation in terms of the Borel parameter $\omega_M$, we obtain
the LO sum rule
\begin{align}
 e^{-m_\Lambda^2/(n_+p') \omega_M}  \, (n_+p') \, f_\Lambda \, \xi_\Lambda(n_+p') 
&= f_{\Lambda_b}^{(2)} \int_0^{\omega_s} d\omega \, \phi_4(\omega) \, e^{-\omega/\omega_M} \,,
\label{xi:LOexact}
\end{align}
which takes the analogous form as for the $B \to \pi, \rho$ case, only that the distribution
amplitude for the spectator anti-quark in the $B$-meson is replaced by the effective LCDA for
the spectator di-quark in the $\Lambda_b$ baryon.

The formal scaling of the (tree-level) result for $\xi_\Lambda$ with the  large-energy variable
can be derived by further considering the limit $\omega_{s,M} \sim \frac{\Lambda_{\rm had.}^2}{n_+p'} 
 \ll \langle \omega \rangle$, which allows one
to expand the LCDA of the $\Lambda_b$ baryon around $\omega=0$ in the integrand. This yields
\begin{align}
 \xi_\Lambda(n_+p') & \approx \frac{f_{\Lambda_b}^{(2)} \, \omega_M^2 \, \phi_4'(0)}{(n_+p') \, f_\Lambda} \, e^{m_\Lambda^2/(n_+p') \omega_M} 
 \left( 1 - e^{-\omega_s/\omega_M} \, (1+\frac{\omega_s}{\omega_M}) \right)  \,.
\label{xi:LOapprox}
\end{align}
where $\phi_4'(0) \sim 1/\omega_0^2$ with $\omega_0 \sim \langle \omega \rangle $ being the typical 
light-come momentum of the light di-quark in the heavy baryon (see Appendix~\ref{app:DAs}). In this limit, the soft 
$\Lambda_b \to \Lambda$ form factor thus scales as $1/(n_+p')^3$ with the large energy of the
final state baryon. Compared to the mesonic case \cite{DeFazio:2005dx}, 
one encounters an additional factor of $1/(n_+p')$ which physically can be traced back to the
phase-space suppression of the additional spectator quark. Technically, the difference between the mesonic
and baryonic case stems from the fact that the B-meson LCDA $\phi_B^-(\omega)$ does not vanish at the endpoint,
while $\phi_4(\omega)$ vanishes linearly.

We should stress that radiative corrections to the sum rule will lead to additional non-analytical 
dependence of the form factors on $(n_+p')$ with logarithmically enhanced perturbative coefficients.
Part of these corrections are universal and can be uniquely factorized in terms of: (i) hard vertex
corrections absorbed in Wilson coefficients of SCET decay currents, (ii) a jet function, absorbing
the hard-collinear emissions from the strange-quark propagator in SCET, (iii) the soft evolution
of the LCDAs of the $\Lambda_b$ baryon. To this accuracy, we obtain
an analogous result as discussed for the mesonic case \cite{DeFazio:2005dx},
\begin{align}
 F_i(q^2) &\simeq C_i(n_+p',\mu) \cdot
\frac{e^{m_\Lambda^2/(n_+p')\omega_M} \, f_{\Lambda_b}^{(2)}}{(n_+p') \, f_\Lambda}
  \int_0^{\omega_s} d\omega' \, e^{-\omega'/\omega_M} 
\cr 
& \quad \Bigg\{ \left[ 1 +  \frac{\alpha_s C_F}{4\pi} 
 \left(
 7-\pi^2
 + 3 \ln \left[\frac{\mu^2}{\omega'(n_+p')}\right]
 + 2 \ln^2 \left[\frac{\mu^2}{\omega'(n_+p')}\right]
\right) \right] \phi_4(\omega',\mu) 
\cr &
 \quad 
{} + \frac{\alpha_s C_F}{4\pi} \, \int\limits_0^{\omega'} d\omega
 \left( 4 \ln \left[\frac{\mu^2}{(\omega'-\omega) (n_+p')}\right] + 3 \right)
 \, \frac{\phi_4(\omega',\mu)-\phi_4(\omega,\mu)}{\omega'-\omega}
\Bigg\}
\,,
\label{LL}
\end{align}
where $F_i(q^2)$ denotes a generic form factor with the corresponding Wilson coefficient $C_i$.
The leading (double-logarithmic) $\mu$-dependence cancels between the 3 terms on the right-hand side, thanks to
the renormalization-group equations (see e.g.\ \cite{Bauer:2000yr,Ball:2008fw,Lange:2003ff,Bell:2008er,DescotesGenon:2009hk}),
\begin{align}
 \frac{d}{d \ln \mu} \, C_i(n_+p',\mu) &=- \frac{\alpha_s C_F}{4\pi}
 \, \Gamma_{\rm cusp}^{(1)} \, \ln \frac{\mu}{M_b} \, C_i(n_+p',\mu) + \ldots
\,,
\\
 \frac{d}{d \ln \mu} \, \phi_4(\omega,\mu) &=- \frac{\alpha_s C_F}{4\pi}
 \, \Gamma_{\rm cusp}^{(1)} \, \ln \frac{\mu}{\omega} \, \phi_4(\omega,\mu) + \ldots   
\end{align}
with the cusp-anomalous dimension $\Gamma_{\rm cusp}^{(1)}=4$.
Evaluating the terms in curly brackets in (\ref{LL}) at a factorization scale of order 
$\mu^2\sim \omega_s (n_+p')$ and evolving the Wilson coefficients down to
that scale, one achieves the resummation of the leading Sudakov double logarithms.

Additional \emph{process-dependent} corrections to (\ref{LL}) arise from hard-collinear gluon exchange between the
strange quark and the spectator quarks in SCET. As shown in \cite{DeFazio:2005dx}, these
will lead to logarithmically enhanced terms which are sensitive to the endpoint behaviour of $\phi_4(\omega,\mu)$.
The explicit derivation of these terms is left for future work.

\subsection{Corrections from Hard-Collinear Gluon Exchange}

As explained above, sub-leading currents in the SCET Lagrangian will induce violations
of the form-factor symmetry relations in the large recoil limit. Contributions
involving hard-collinear gluon exchange can be treated perturbatively
in SCET correlation functions. The leading effect requires one to calculate the matrix
element in (\ref{deltasoft}),
whose leading contribution arises from hard-collinear gluon exchange with
one of the two spectator quarks in the baryons, see the corresponding diagram
on the r.h.s.\ of Fig.~\ref{fig}. From the perspective of QCD factorization,
this diagram represents an intermediate (hybrid) case, where some of the constituents
undergo calculable short-distance interactions, while the remaining spectator quark remains
undisturbed and is thus forced to populate the endpoint region in phase space.

In the sum-rule approach, as before, we define a correlation function (in light-cone gauge)
\begin{align}
 \Pi_\Lambda^\mu(n_-p') & \equiv i \int d^4x \, e^{ip'x} \langle 0| T \left[ \frac{\slash n_+ \slash n_-}{4} \, J_\Lambda(x) \left[ \bar s(0) \, \tilde \Gamma 
  \, g A^\mu_\perp(0) \, b(0) \right] \right] |\Lambda_b(p)\rangle \,.
\end{align}
Notice that this time, we have to use the opposite light-cone projector acting on $J_\Lambda$, as compared 
to the correlation function used to extract the soft form factor $\xi_\Lambda$. It projects on the sub-leading
transverse momentum in the numerator of the strange-quark propagator which is required from rotational invariance in the transverse plane.
The light-quark momenta in the $\Lambda_b$ baryon again are denoted as $k_{1,2}$ respectively, with $k_1 = \omega_1+k_{1\perp}$, 
$k_2 = \omega_2+k_{2\perp}$, and $\omega=\omega_1+\omega_2 =n_-k$. Using the momentum-space projector for the LCDAs of $\Lambda_b$ as 
given in Appendix~\ref{app:DAs},
and assuming isospin symmetry of strong interactions, we obtain
\begin{align}
& \Pi^\mu(n_-p') = -i \, g_s^2 \, C_F \, \frac{f_{\Lambda_b}^{(2)}}{4}  \, 
 \int d\omega_1 \int d\omega_2 \cr 
 & \quad \times \int \frac{d^Dl}{(2\pi)^D} \, 
\frac{1}
{[l_\perp^2 + (n_+l)(n_-l)] 
 [l_\perp^2+(n_+l) \, (n_-l-\omega_2)] 
 [l_\perp^2+(n_+p'+n_+l)(n_-p'+ n_-l -\omega)]}
\cr & \quad \times 
{\rm tr}\left[\tilde M(k_1,k_2) \, C\gamma_5 \slash n_+ (\slash k_2-\slash l) \gamma^\mu_\perp \right]
\frac{\slash n_+ \slash n_-}{4} \, (\slash l-\slash k_1-\slash k_2) \, \tilde\Gamma \, u_{\Lambda_b}(v,s) \, ,
\end{align}
\noindent
where the square bracket around propagator denominators imply a ``$+i\epsilon$'' description. 
The Dirac trace is easily calculated as 

\begin{align}
 & {\rm tr}\left[\tilde M(k_1,k_2) \, C\gamma_5 \slash n_+ (\slash k_2-\slash l) \gamma^\mu_\perp \right] 
\cr 
& \quad = - 4 \psi_4(\omega_1,\omega_2)\, l_\perp^\mu 
   + 2 (n_+l)  \left( G(\omega_1,\omega_2) \,
 \frac{\partial}{\partial k_{1\mu}^\perp}  +  H(\omega_1,\omega_2) \,
 \frac{\partial}{\partial k_{2\mu}^\perp} \right) \,.
\end{align}

\noindent This yields
\begin{align}
 & \Pi^\mu(n_-p') = i \, \frac{ g_s^2 C_F f_{\Lambda_b}^{(2)}}{2}  \, 
 \int d\omega_1 \int d\omega_2 \cr 
 & \quad \times \int \frac{d^Dl}{(2\pi)^D} \, \frac{2l_\perp^2/(D-2) \, \psi_4(\omega_1,\omega_2) +
 (n_+l)  \left(G(\omega_1,\omega_2)  + H(\omega_1,\omega_2)\right)}
{[l_\perp^2 + (n_+l)(n_-l)] 
 [l_\perp^2+(n_+l) \, (n_-l-\omega_2)] 
 [l_\perp^2+(n_+p'+n_+l)(n_-p'+ n_-l -\omega)]}
\cr & \quad \times 
\frac{\slash n_+ \slash n_-}{4} \, \gamma^\mu_\perp \, \tilde\Gamma \, u_{\Lambda_b}(v,s) \,.
\end{align}

\noindent Notice that both terms contribute at the same order in the SCET correlator, 
since $l_\perp^2 \sim (n_+l) \omega \sim m_b \Lambda$. However, the contributions from
$\psi_4$ and $G$ will give formerly sub-leading contributions to the sum-rule for $\omega_1\to 0$. 
Performing the integration over $(n_-l)$ and $l_\perp$, the Borel transformation and continuum subtraction,
we obtain
\begin{align}
 \hat B \Pi_\Lambda^\mu(\omega_M)|_{\rm subtr.} &=
- \frac{ \alpha_s C_F f_{\Lambda_b}^{(2)}}{4\pi}  \, 
 \int d\omega_1 \int d\omega_2 \int_0^{\omega_s}
 \frac{d\omega'}{\omega_M} \, e^{-\omega'/\omega_M} \cr 
 & \quad \times  \left\{ 
    \frac{ \left(\omega_2+(\omega'-\omega) \theta(\omega-\omega') \right) 
  \theta(\omega'-\omega_1)}{4 \omega_2} 
  \, \psi_4(\omega_1,\omega_2) \right.
\cr
& \qquad \left.
 + \frac{\theta(\omega-\omega') \theta(\omega'-\omega_1)}{2\omega_2}  \left(G(\omega_1,\omega_2)  + H(\omega_1,\omega_2)\right)
\right\}
\cr & \quad \times 
\frac{\slash n_+ \slash n_-}{4} \, \gamma^\mu_\perp \, \tilde\Gamma \, u_{\Lambda_b}(v,s) \,,
\end{align}

In the limit $\omega_s,\omega_M \ll \langle\omega_{1,2}\rangle$, the typical momentum of the light quarks in the heavy baryon,
the integral can be simplified. Since $\omega_1 \leq \omega' \leq \omega_s$, we may approximate $\omega_1 \simeq 0$ in the LCDAs. 
This reflects the fact that now, the hard-collinear scattering requires the struck spectator-quark to carry almost all the momentum 
$\omega$ of the di-quark compound. In this limit, we have
\begin{align}
\hat B \Pi_\Lambda^\mu(\omega_M)|_{\rm subtr.} &\simeq
- \frac{\alpha_s C_F}{8\pi}\, \gamma^\mu_\perp \, \tilde \Gamma  \, u_{\Lambda_{b}}(v,s) 
 \underbrace{f_{\Lambda_{b}}^{(2)} \, \int_0^\infty \! \frac{d\omega}{\omega} \, H(0,\omega)} \times 
 \underbrace{\left(\omega_M - e^{-\omega_s/\omega_M} (\omega_M+\omega_s) \right)}
\,,
\cr 
 & \hskip13.6em \Lambda_b \hskip8.2em \qquad J_\Lambda
\label{delxilim}
\end{align}

\noindent and the correlation function factorizes, as indicated, into an inverse moment of the heavy-baryon LCDA and
a function of the Borel and threshold parameter describing the spectrum of the interpolating current for
the light baryon.
For the hadronic side of the sum rule, the contribution of the $\Lambda$ baryon to the correlator
is now given by
\begin{align}
 \Pi_\Lambda^\mu \big|_{\rm reson.} 
&= \frac{f_\Lambda \, m_\Lambda \, M_{\Lambda_b} \, \Delta\xi_\Lambda}{m_\Lambda^2/(n_+p')- (n_-p')} \, \gamma^\mu_\perp \, \tilde \Gamma  \, u_{\Lambda_b}(v,s)  \,,
\end{align}
which leads to the sum rule
\begin{align}
 & e^{-m_\Lambda^2/(n_+p')\omega_M} \, f_\Lambda \, M_{\Lambda_b} \, m_\Lambda/\omega_M \, \Delta \xi_\Lambda \cr
 & \quad \quad = - \frac{ \alpha_s C_F f_{\Lambda_b}^{(2)}}{4\pi}  \, 
 \int d\omega_1 \int d\omega_2 \int_0^{\omega_s}
 \frac{d\omega'}{\omega_M} \, e^{-\omega'/\omega_M} \cr
 & \quad \quad \quad \times \left\{\frac{ \left(\omega_2+(\omega'-\omega) \theta(\omega-\omega') \right) 
  \theta(\omega'-\omega_1)}{4 \omega_2} \psi_4(\omega_1,\omega_2) \right. \cr 
 & \quad \quad \quad \quad  \left. + \frac{\theta(\omega-\omega') \theta(\omega'-\omega_1)}{2\omega_2} \left(G(\omega_1,\omega_2) + H(\omega_1,\omega_2)\right)
\!\right\}
\label{eq:delxi}
\\
& \quad \quad \simeq - \frac{\alpha_s C_F}{8\pi} \, f_{\Lambda_{b}}^{(2)} \, \int_0^\infty \! \frac{d\omega}{\omega} \, F(0,\omega) \times 
 \left(\omega_M - e^{-\omega_s/\omega_M} (\omega_M+\omega_s) \right) \,.
\label{eq:delxilimit}
\end{align}

The correction to the soft form factor, in the large recoil limit, thus scales as
$$
  \Delta \xi_\Lambda/\xi_\Lambda \sim \alpha_s \, \frac{\omega_0}{m_\Lambda} \, \frac{n_+p'}{M_{\Lambda_b}} \,,  
$$
i.e.\ it formally has the same power-counting in terms of $\Lambda_{\rm QCD}/M_b$ (although, numerically,
the ratio $\omega_0/m_\Lambda$ is small), but a less
pronounced $(n_+p')$ dependence than the soft form factor. Notice that in the ratio $\Delta \xi_\Lambda/\xi_\Lambda$,
the dependence on the baryon decay constants drops out, while the sensitivity to the sum-rule
parameters and the features of the LCDAs of the $\Lambda_b$ baryon remains.


\section{Numerical Results}

\label{sec:numerics}

\begin{table}[b]
\caption{\small \label{tab:input} \small Summary of hadronic input parameters}
 \begin{center}
   \begin{tabular}{l| c | r}
    \hline
     Parameter & central value & remarks
\\
 \hline \hline
 threshold $s_0$ & 2.55~GeV$^2$    & $\Lambda(1600)$ 
\\
$\omega_s \equiv s_0/(n_+p')$ &
\\[0.2em]
\hline
 Borel $M_{\rm Borel}^2$ & 2.5~GeV$^2$ &  
\\
$\omega_M \equiv M_{\rm Borel}^2/(n_+p')$  &
\\[0.2em]
\hline
 decay constant $f_\Lambda$ & $6 \cdot 10^{-3}$~GeV$^2$  & \cite{Liu:2008yg}
\\
\hline
 decay constant $f_{\Lambda_b}^{(2)}$ & $0.030$~GeV$^3$  & \cite{Ball:2008fw}
\\
 LCDA par.\ $\omega_0$ & 300~MeV & (our estimate)
\\
\hline 
   \end{tabular}

 \end{center}

\end{table}

In the following section we present some numerical results for the soft $\Lambda_b\to\Lambda$ form factor $\xi_\Lambda$
and the correction from hard-collinear gluon exchange, $\Delta \xi_\Lambda$, in the large-recoil limit.
The numerical predictions involve a number of hadronic parameters with respective uncertainties, 
for which we summarize our default choices in Table~\ref{tab:input} for convenience. For the shape of the LCDAs,
we use the simple exponential models as summarized in Appendix~\ref{app:DA1}.

\subsection{Soft Form Factor}

\begin{figure}[t!pbt]

 \begin{center}
  \includegraphics[height=0.29\textwidth]{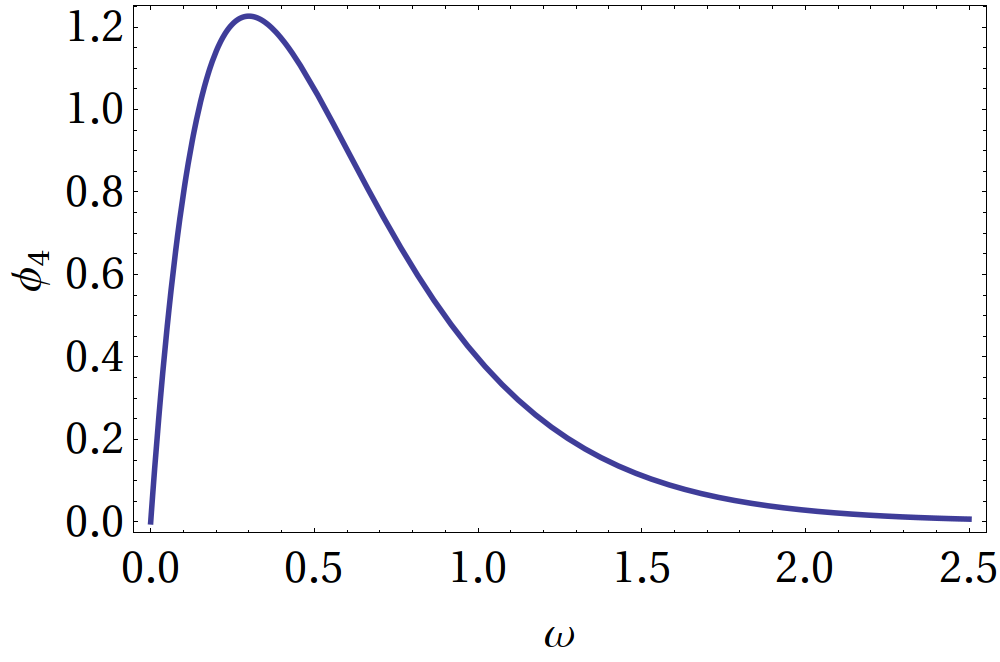}
\qquad
  \includegraphics[height=0.29\textwidth]{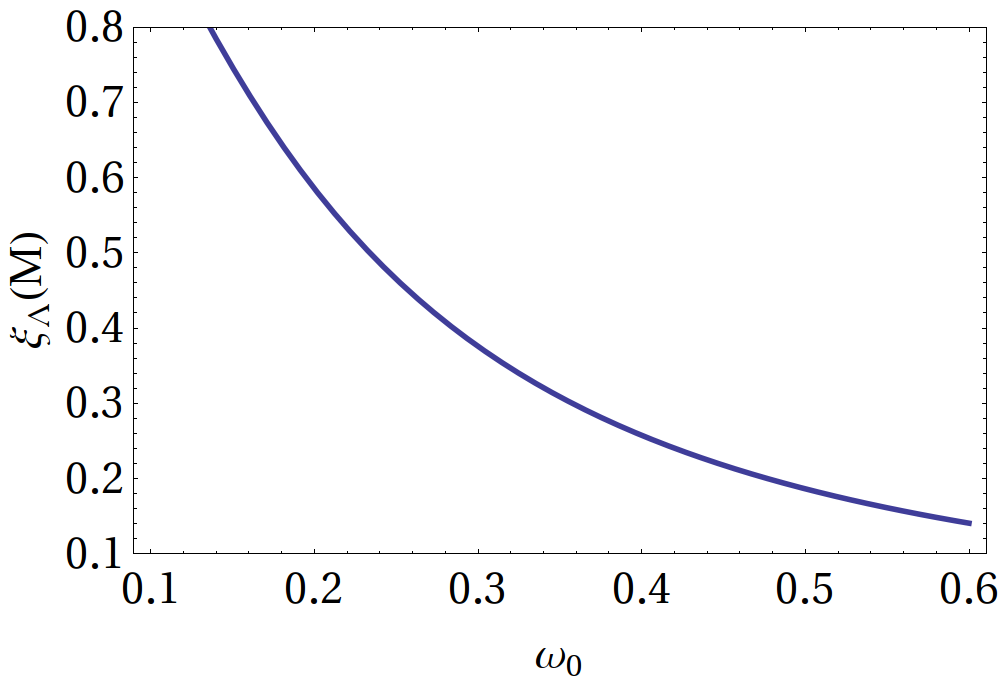}
\caption{\small \label{fig:phi4} \small Left: Functional form of the partially integrated LCDA $\phi_4(\omega)$
  for the exponential model and $\omega_0=300$~MeV.
  Right: Dependence of $\xi_\Lambda(n_+p'=M_{\Lambda_b})$ on the value of $\omega_0$.}
\end{center}
\end{figure}

The value for the soft form factor is estimated from the LO sum rule (\ref{xi:LOexact}).
We will also compare with the approximation (\ref{xi:LOapprox}).
The default value for the threshold parameter 
is taken from the position of the next highest $b$-baryon resonance\footnote{One should, however, be aware that 
one may encounter pollution from baryon states with opposite parity, see
the recent discussion in \cite{Khodjamirian:2011jp}.}
with $I(J^P)=0(1/2^+)$.
For the relevant LCDAs, we will use the model (\ref{Lamb:LCDA})
as described in the Appendix. In the soft form factor, only the partially integrated
function $\phi_4(\omega)$ appears. In our model, it takes the simple form
$$
  \phi_4(\omega) := \frac{\omega}{\omega_0^2} \, e^{-\omega/\omega_0} \,,
$$
which is illustrated on the left of Fig.~\ref{fig:phi4}.

For the default parameter values in Table~\ref{tab:input}, 
the soft form factor at maximal recoil is estimated as
$$
  \xi_\Lambda(n_+p'=M_{\Lambda_b}) \simeq 0.38 \qquad \mbox{central value, from (\ref{xi:LOexact}),}
$$
which -- within the uncertainties -- is consistent with estimates from other methods in \cite{Aliev:2010uy,Chen:2001zc}.
We remark in passing, that the authors 
of \cite{Wang:2009hra} estimate the $\Lambda_b \to \Lambda$ form factors with
a similar set-up, but without performing the large-recoil limit in SCET explicitly. They
quote a rather small value $g_2(q^2=0) = 0.018 \pm 0.003$ for one of the form factors that,
as we understand, should coincide with $\xi_\Lambda(n_+p'=M_{\Lambda_b})$ in the heavy-quark limit.

\begin{figure}[p!tpb]

 \begin{center}
  \includegraphics[width=0.46\textwidth]{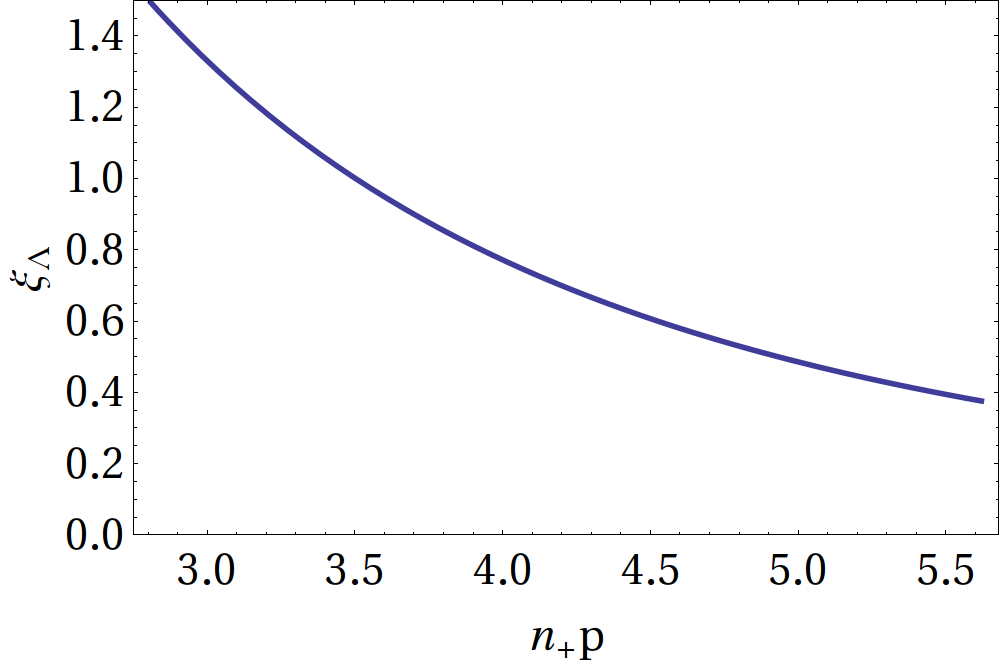}
\qquad
  \includegraphics[width=0.46\textwidth]{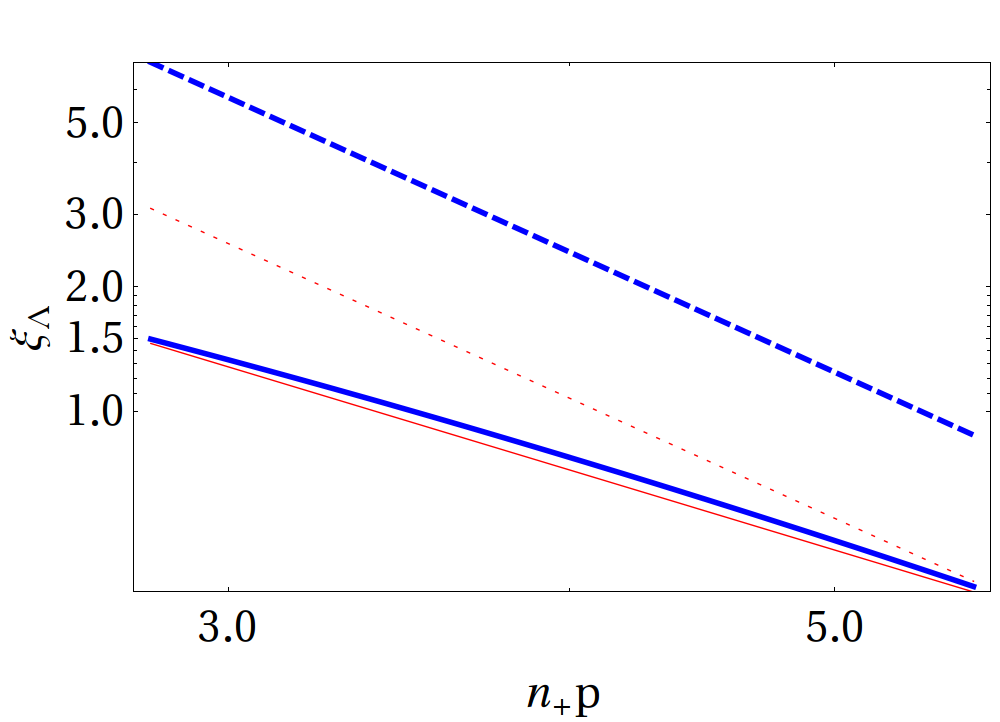}
\caption{\small \label{fig:npp} 
  Dependence of the soft form factor on $(n_+p')$. Left: energy dependence 
  from the LO sum rule (\ref{xi:LOexact}). Right: comparison of the 
  LO sum rule (\ref{xi:LOexact}) --thick solid line -- , with 
  the approximate formula (\ref{xi:LOapprox}) -- thick dashed line --, 
  and a power-like behaviour with $1/(n_+p')^3$ (dash-dotted) or $1/(n_+p')^2$ (dotted).}
\end{center}
\end{figure}

\begin{figure}[h!tpb]

 \begin{center}
  \includegraphics[height=0.29\textwidth]{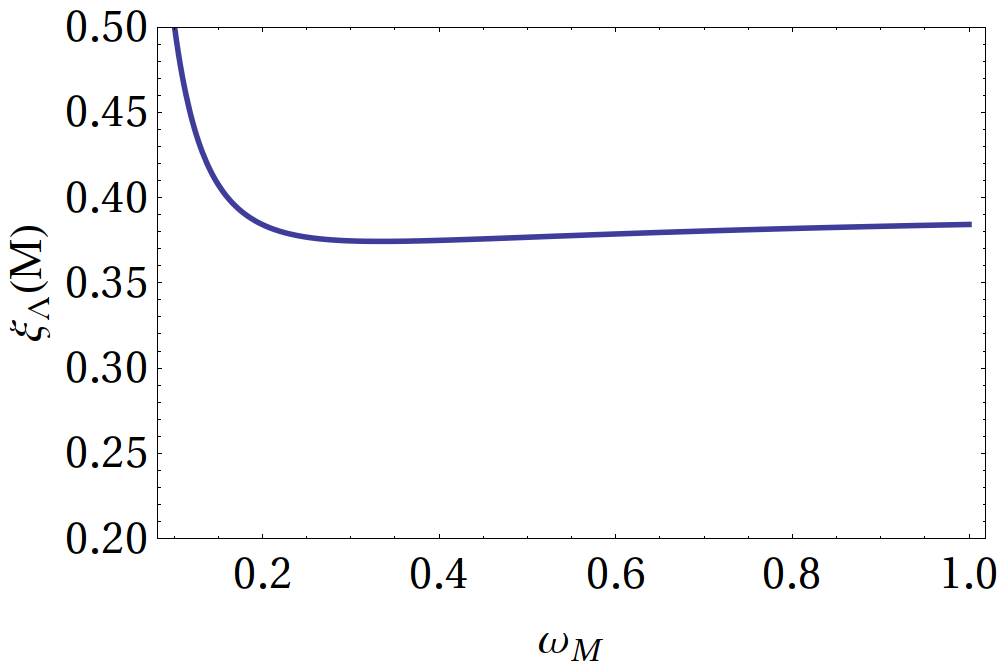}
\qquad
  \includegraphics[height=0.29\textwidth]{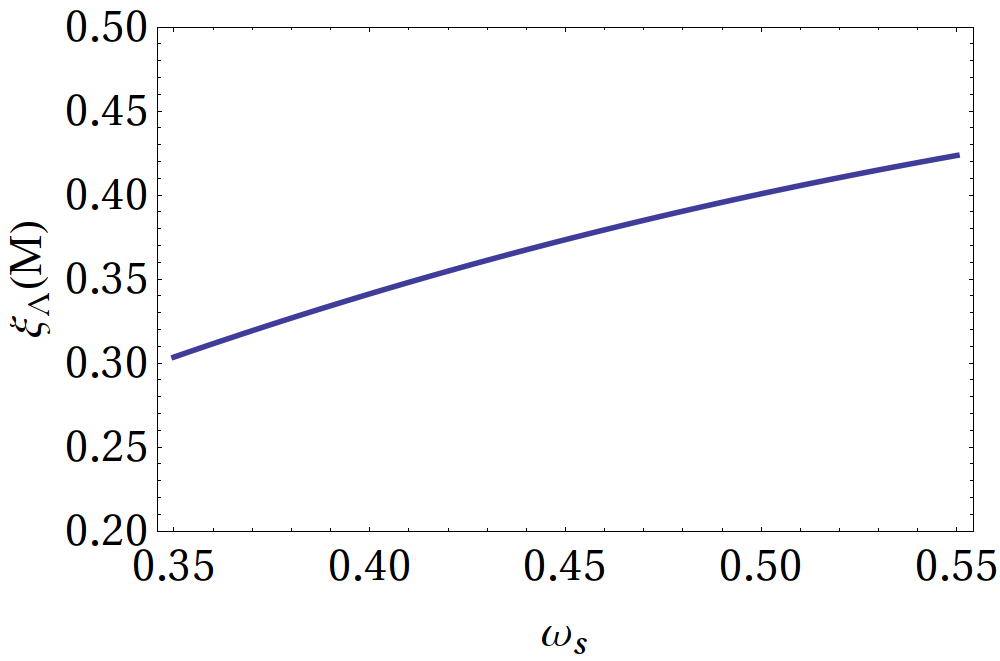}
\caption{\small \label{fig:wswm} \small 
  Dependence of the soft form factor on the sum-rule parameters (for maximal recoil, $n_+p'=M_{\Lambda_b}$). 
  Left: Dependence on the Borel parameter $\omega_M$. Right: Dependence on
    the threshold parameter $\omega_s$.}
\end{center}
\end{figure}

The dependence of $\xi_\Lambda$ on the LCDA parameter $\omega_0$ is shown on the
right of Fig.~\ref{fig:phi4}.
The energy dependence is plotted in Fig.~\ref{fig:npp}.
The dependence on the sum-rule parameters (at maximal recoil) is shown in
Fig.~\ref{fig:wswm}.
The following observations can be made:
\begin{itemize}
 \item For values of $\omega_0$ around 300~MeV or smaller, as extracted from the analysis
     in \cite{Ball:2008fw}, the approximate formula (\ref{xi:LOapprox})
    does not yield a reliable estimate, because 
    numerically $\omega_0 \simeq \omega_s \simeq \omega_M$. The respective value of $\xi_\Lambda$
    is overestimated by more than a factor 2 in this case. On the other hand, 
    compared to the mesonic case, one might have expected larger values of $\omega_0$ in
    the baryonic LCDA in the first place.

 \item In any case, the sum-rule result for $\xi_\Lambda$ is very sensitive to the 
    shape of the LCDA in general and the value
    of $\omega_0$ in particular. 
Varying $\omega_0$ in a reasonable range between $0.2$ and $0.5$~GeV, induces
a 50\%\ uncertainty on $\xi_\Lambda$.
More independent information on the LCDAs of the $\Lambda_b$ baryon and the
relevant hadronic parameters is clearly 
needed to reach reasonable precision in this kind of sum-rule analysis.

 \item For small values of $\omega_0$, the energy dependence of the form factor
    follows an approximate $1/(n_+p')^2$ behaviour, rather than a $1/(n_+p')^3$
    behaviour as predicted by (\ref{xi:LOapprox}).

 \item The dependence on the Borel parameter is very weak (less than a few percent) 
    and negligible compared to the other uncertainties.

 \item The dependence on the threshold parameter is almost linear, and the LO sum-rule
    result thus depends on the modelling of the continuum contribution to the
    correlator in an essential way. Varying $\omega_s$ between $0.35$ and $0.55$~GeV,
    the induced uncertainty for $\xi_\Lambda$ at maximal recoil amounts to about 10-20\%.

\end{itemize}


Taking these observations at face value, we have to conclude
that the normalization of the $\Lambda_b \to \Lambda$ form factors
at large recoil still suffers from sizeable uncertainties, mostly from
the $\Lambda_b$ LCDAs and the threshold parameter. The same is true
for the energy-dependence of the form factor which varies between
a $1/(n_+p')^2$ behaviour (small values of $\omega_0$) and
a $1/(n_+p')^3$ behaviour (large values of $\omega_0$). Independent
information on the LCDA $\phi_4(\omega)$ and/or on the $\Lambda_b \to
\Lambda$ form factors at intermediate momentum transfer 
from Lattice QCD would clearly be helpful in this context.


\subsection{Form-Factor Ratios}

\begin{figure}[tpb]
 \begin{center}
  \includegraphics[height=0.29\textwidth]{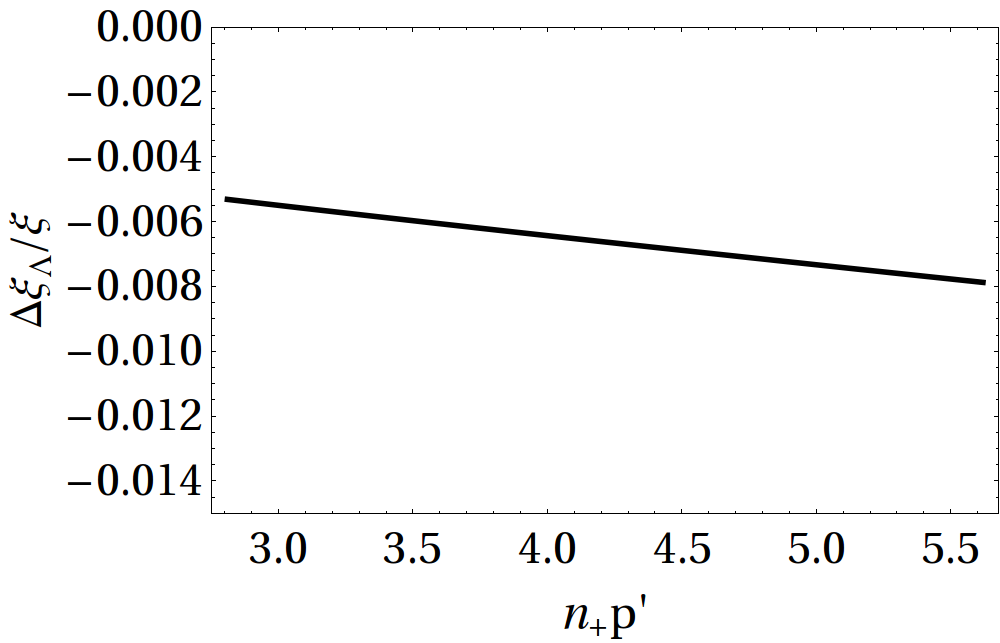}
\qquad
  \includegraphics[height=0.29\textwidth]{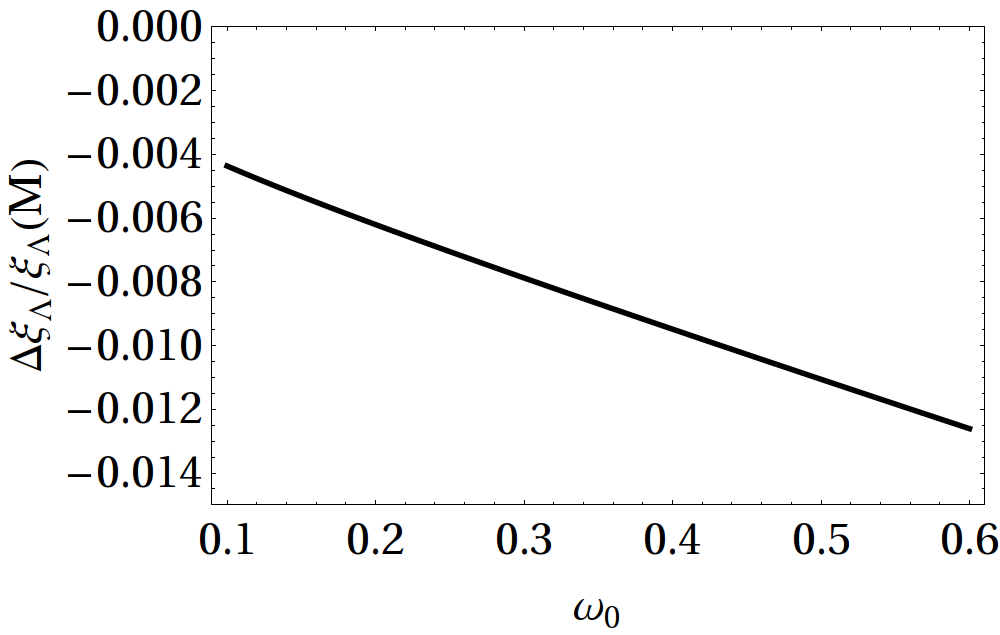}
\caption{\small \label{fig:delxi1} 
  Form-factor correction $\Delta \xi_\Lambda/\xi_\Lambda$ from the exchange of one hard-collinear
  gluon from SCET sum rules. Left: Energy dependence 
  from the LO sum rules (\ref{xi:LOexact}, \ref{eq:delxi}). Right: Dependence
  on the parameter $\omega_0$ characterizing the LCDAs of the $\Lambda_b$ baryon.}
\end{center}
\end{figure}

The symmetry relations between the individual $\Lambda_b \to \Lambda$ form factors
receive perturbative and non-perturbative corrections. Let us first consider
the corrections from the exchange of one hard-collinear gluon, contributing to
the function $\Delta \xi_\Lambda$ as estimated from the sum rule in (\ref{eq:delxi}). 
For the default values of the hadronic input parameters, we take the same
values as before, see Table~\ref{tab:input}. As the default value for the
strong coupling constant at a hard-collinear scale, we use $\alpha_s \simeq \alpha_s(\mu=2~{\rm GeV}) \simeq 0.3$.
For the relevant LCDAs, we will again use the exponential model discussed in section \ref{app:DA1}. 
With this, we obtain as our default estimate
$$
  \Delta\xi_\Lambda(n_+p'=M_{\Lambda_b}) \simeq - 0.003 \,, \qquad 
\frac{\Delta\xi_\Lambda}{\xi_\Lambda} \simeq -0.8\% \,.
$$
We also find that the ratio $\Delta \xi_\Lambda/\xi_\Lambda$ exhibits a mild linear dependence on
the (large) recoil-energy and a pronounced linear dependence on the parameter $\omega_0$ in
the exponential model for the $\Lambda_b$ LCDAs, see Fig.~\ref{fig:delxi1}. 
This is in qualitative agreement with the considerations after Eq.~(\ref{eq:delxilimit}).

\begin{figure}[tpb]
 \begin{center}
  \includegraphics[height=0.29\textwidth]{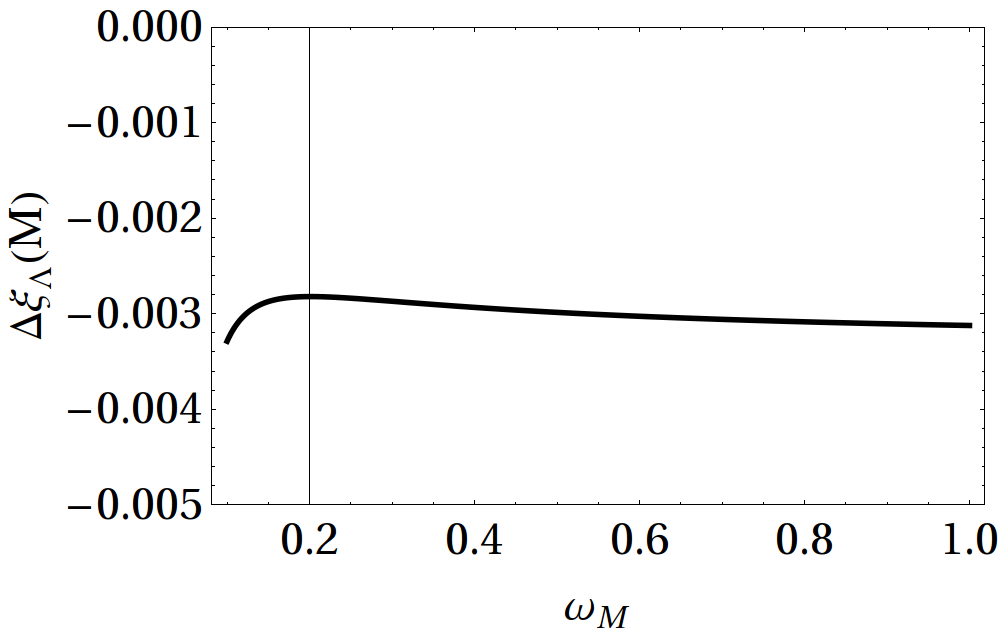}
\quad
  \includegraphics[height=0.29\textwidth]{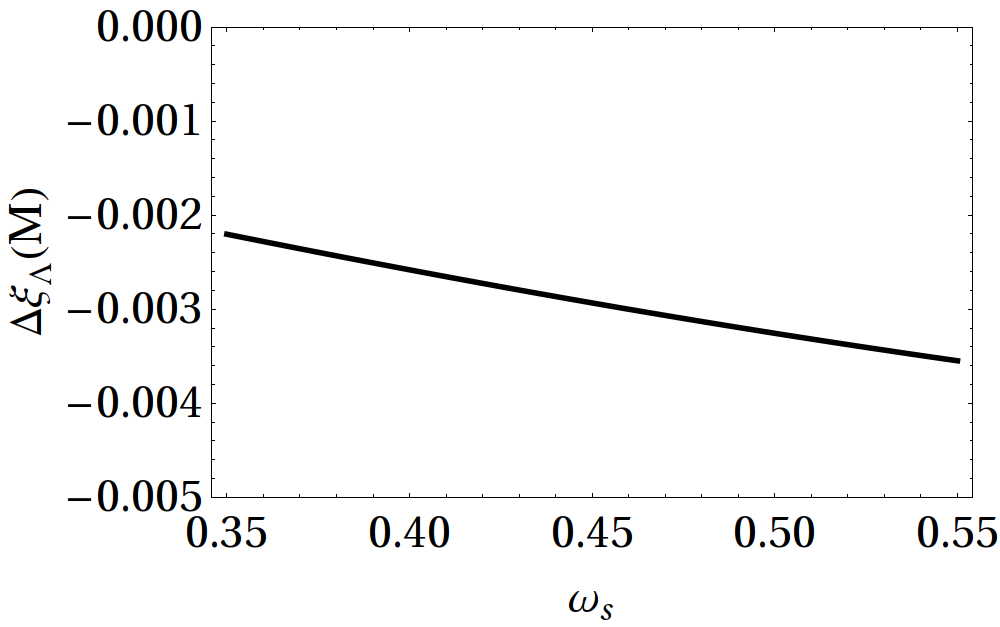}
\caption{\small \label{fig:delxi2} 
  Dependence on $\Delta \xi_\Lambda$ on the sum-rule parameters (for $n_+p'=M_{\Lambda_b}$). 
  Left: Dependence on the Borel parameter $\omega_M$. Right: Dependence on
    the threshold parameter $\omega_s$.}
\end{center}
\end{figure}

The dependence of $\Delta \xi_\Lambda$ on the sum-rule parameters is plotted in
Fig.~\ref{fig:delxi2}. The sensitivity to the Borel parameter $\omega_M$, again, 
is rather weak, while the dependence on the threshold parameter $\omega_s$ is
somewhat weaker than for the soft form factor $\xi_\Lambda$. Because of the different systematics
in (\ref{xi:LOexact}, \ref{eq:delxi}) related to the modelling of the continuum and the pollution from other baryonic
resonances, the dependence of the ratio $\Delta \xi_\Lambda/\xi_\Lambda$
on the sum rule parameters is difficult to estimate numerically.
As already emphasized, the dependence on the light and heavy decay constants drops out
in the ratio $\Delta \xi_\Lambda/\xi_\Lambda$. The overall dependence on the 
renormalization scale used for the strong coupling constant has to be resolved by
calculating higher-order radiative corrections to $\Delta \xi_\Lambda$ in SCET.


\begin{figure}[tpb]
 \begin{center}
  \includegraphics[height=0.29\textwidth]{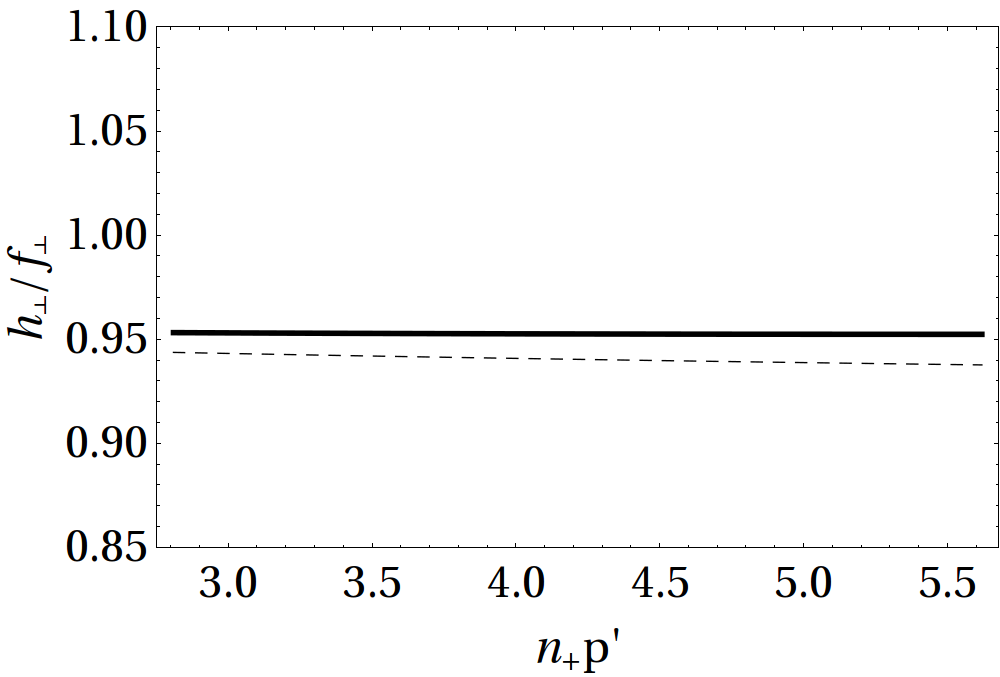}
\qquad
  \includegraphics[height=0.29\textwidth]{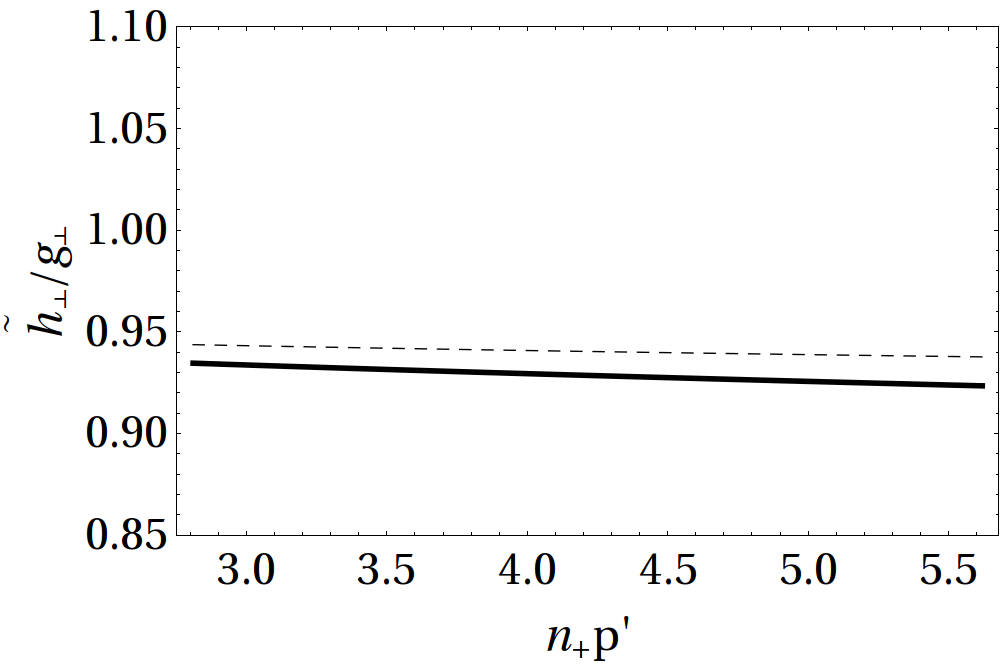}
\caption{\small \label{fig:ffratios} 
  Form-factor ratios including ${\cal O}(\alpha_s)$ corrections
  from hard (dashed line) and hard plus hard-collinear (solid line) gluon exchange,
  as a function of the recoil energy $(n_+p')$:
  Left: The ratio $h_\perp/f_\perp$. Right: The ratio $\tilde h_\perp/g_\perp$.}
\end{center}
\end{figure}

The above result can be turned into an estimate for form-factor ratios appearing
in physical decay observables. As an example, we discuss the ratios
$$
\frac{h_\perp}{f_\perp} \,, \qquad
\frac{\tilde h_\perp}{g_\perp} \,,
$$
appearing in the forward-backward asymmetry for $\Lambda_b \to \Lambda \mu^+ \mu^-$, see below. 
Including the effect of hard-vertex corrections to ${\cal O}(\alpha_s)$ accuracy,
we obtain the results shown in Fig.~\ref{fig:ffratios}, where we have used 
$$
\alpha_s(m_b) \simeq 0.2
$$
in the hard vertex corrections. As one can see, the corrections to the form-factor 
ratios are dominated by the hard gluon effects in the matching coefficients for the decay currents.

\subsection{$\Lambda_b \to \Lambda \mu^+\mu^-$ Observables}

\begin{figure}[p!pb]
 \begin{center}
  \includegraphics[height=0.3\textwidth]{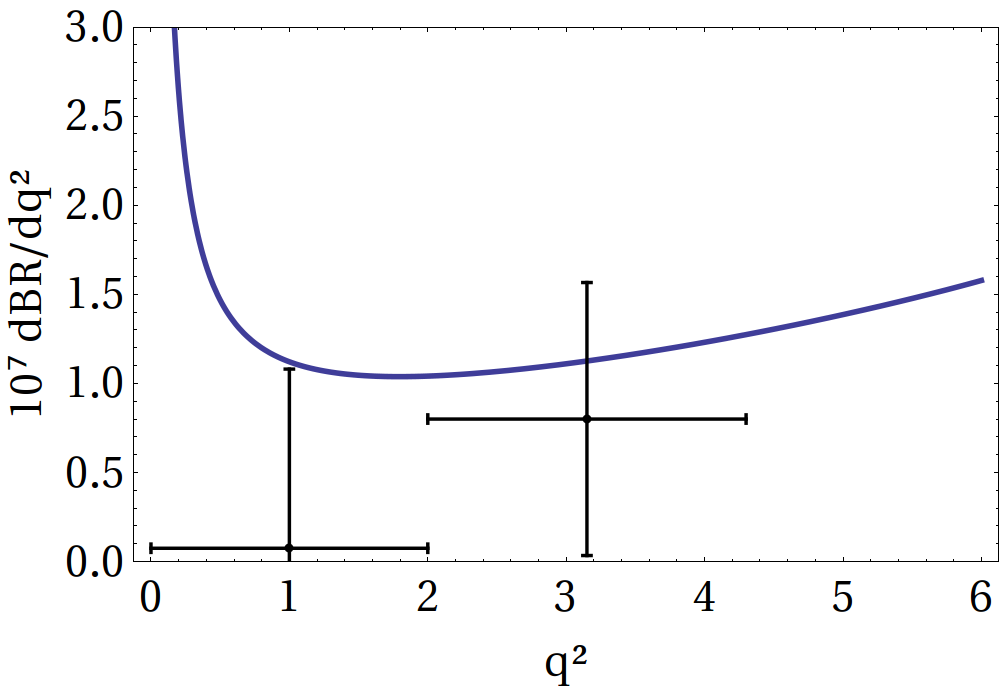}
\caption{\small \label{fig:dBR} 
  Differential branching ratio for $\Lambda_b \to \Lambda \mu^+\mu^-$ in units of $10^{-7}$ as a function of 
 $q^2$ at large recoil. The
theoretical estimate refers to the SCET limit, and the data points are taken from CDF \cite{Aaltonen:2011qs}.
The (substantial) theoretical uncertainties are not shown. }
\end{center}
\end{figure}

The general expressions for the double-differential $\Lambda_b \to \Lambda \mu^+\mu^-$ decay rate
(excluding the non-factorisable contributions, see below) are summarized in Appendix~\ref{app:gammas}.
Our default values for the form-factor estimates, in the large-recoil region,
yield branching ratios which are slightly higher than the central experimental values reported
by CDF \cite{Aaltonen:2011qs} (and compatible with  an independent theoretical estimate
in \cite{Aliev:2010uy}) within the theoretical and experimental uncertainties, see Fig.~\ref{fig:dBR}
(in view of the large hadronic uncertainties, the spectator effects from $\Delta \xi_\Lambda$ represent
a sub-leading effect and are not included here for simplicity).

The functions describing the transverse and longitudinal rate,
and the forward-backward asymmetry  become particularly simple in the SCET limit, where all rates
are proportional to the unique form factor $\xi_\Lambda(n_+p')$, and $m_\Lambda \ll M_{\Lambda_b}$.
To first approximation, the following \emph{ratios} of observables 
are thus independent of hadronic form-factor uncertainties,
\begin{align}
\frac{ H_L(q^2)}{H_T(q^2)} &
\simeq  \frac{q^2}{2 M_{\Lambda_b}^2}\cdot
\frac{ \left|M_{\Lambda_b}^2 \, C_9^{\rm eff}(q^2) 
+ 2 M_b M_{\Lambda_b} \, C_7^{\rm eff} \right|^2
+ \left|M_{\Lambda_b}^2 \, C_{10} \right|^2 }{
\left| q^2 \, C_9^{\rm eff}(q^2)  
+ 2 M_b M_{\Lambda_b} \, C_7^{\rm eff} \right|^2 
+ \left| q^2 \, C_{10} \right|^2 } \,,
\label{r1}
\end{align}
and
\begin{align}
 \frac{H_A(q^2)}{H_T(q^2)}
 &\simeq - 
\frac{ 2 \, {\rm Re} \left[ \left( q^2 \, C_9^{\rm eff}(q^2) 
+ 2 M_b M_{\Lambda_b} \, C_7^{\rm eff}  \right)^*
 q^2 \, C_{10} \right]}{
\left| q^2 \, C_9^{\rm eff}(q^2)  
+ 2 M_b M_{\Lambda_b} \, C_7^{\rm eff} \right|^2 
+ \left| q^2 \, C_{10} \right|^2 
} 
 \,.
\label{r2}
\end{align}
In particular, the leading-order result for the forward-backward asymmetry zero, $q_0^2$,
is determined by the same relation between Wilson coefficients,
\begin{align}
 {\rm Re}\left[ q^2 \, C_9^{\rm eff}(q^2) 
+ 2 M_b M_{\Lambda_b} \, C_7^{\rm eff} \right]_{q^2=q_0^2} & \simeq 0 \,,
\end{align}
as known from the inclusive $b \to s \ell^+\ell^-$ or exclusive $B \to K^*\ell^+\ell^-$ decays
(see \cite{Antonelli:2009ws} and references therein).

\begin{figure}[p!!!pbt]
 \begin{center}
  \includegraphics[height=0.29\textwidth]{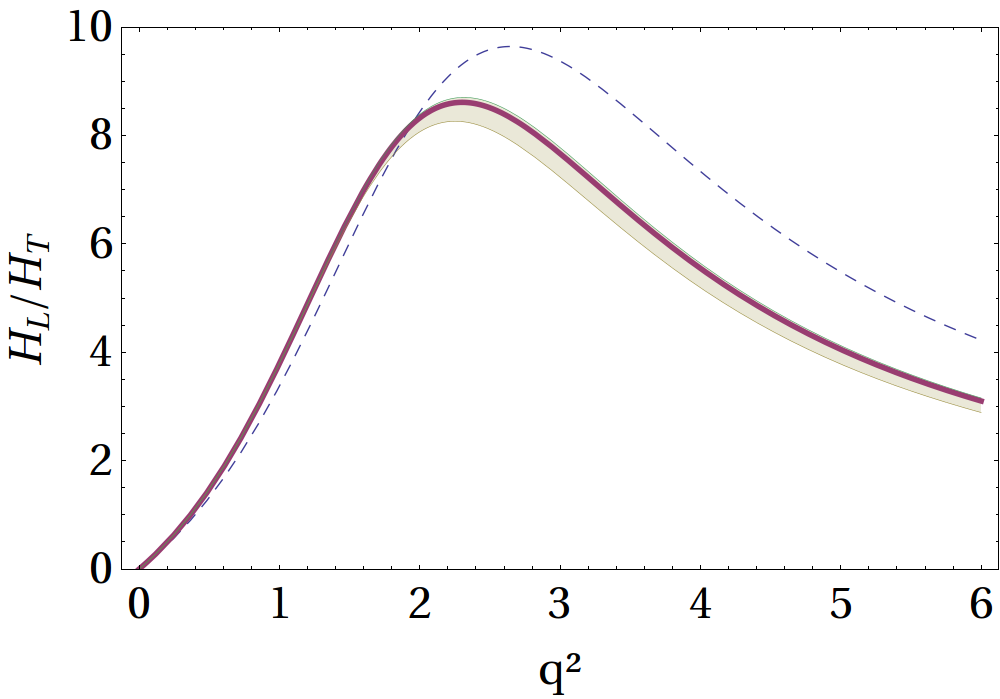}
\qquad
  \includegraphics[height=0.29\textwidth]{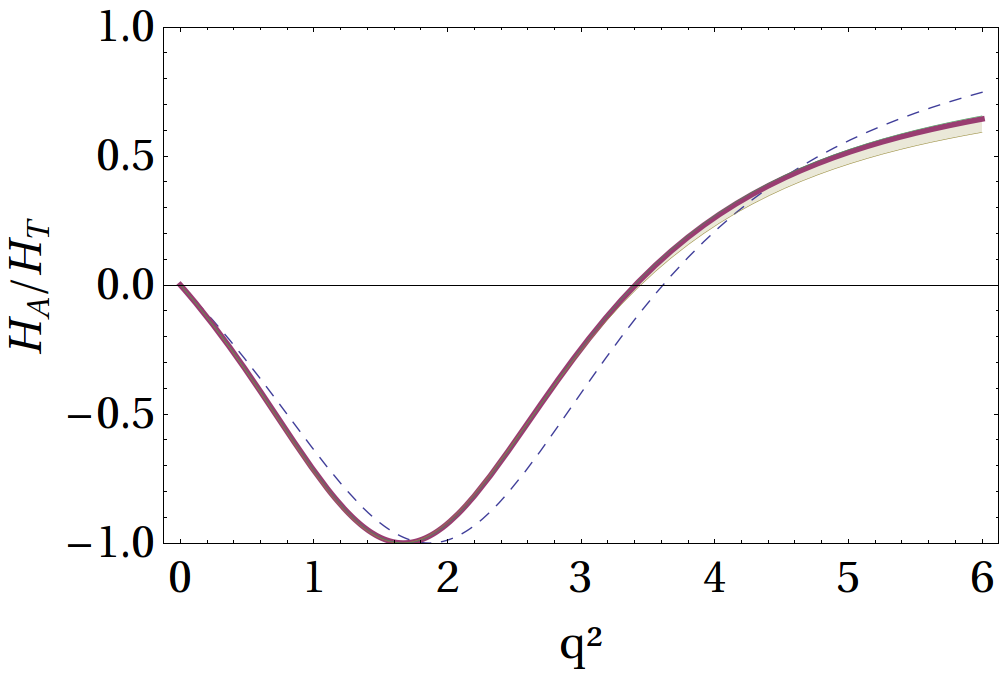}
\caption{\small \label{fig:obsratios} 
  Ratios of observables, $H_L/H_T$ (left) and $H_A/H_T$ (right) as a function of $q^2$.
  The dashed line indicates the SCET limit (\ref{r1}, \ref{r2}). 
  The solid line includes the default estimates for the form-factor corrections from hard gluons, $C_{f_i}$, 
  and hard-collinear gluons, $\Delta \xi_\Lambda$, as well as the kinematic corrections of order $m_\Lambda/M_{\Lambda_b}$. 
  In order to illustrate the (tiny) uncertainty from the variation of
  $\Delta \xi_\Lambda/\xi_\Lambda$, we have inflated the error to an 
  interval $[25\%,400\%]$ of its default value.
 }
\end{center}
\end{figure}

Our numerical estimates for the ratios $H_L/H_T$ and $H_A/H_T$ as a function of $q^2$ are plotted
in Fig.~\ref{fig:obsratios}, where we compare the SCET limit  (\ref{r1},\ref{r2}) with the 
more general result given in (\ref{eqs:obsall}) in the Appendix (which, however, still misses the 
non-factorisable results). In the numerical analysis, the Wilson coefficients $C_{1-7}$ are 
included to leading-logarithmic\footnote{As already mentioned, a complete next-to-leading order analysis
would require one to take into account the non-factorisable gluon corrections, which is left for future work.}
accuracy, and the Wilson coefficients $C_{9,10}$ to next-to-leading
logarithmic accuracy, with the numerical values taken from the analysis in \cite{Beneke:2004dp}.
As one can observe, the inclusion of the kinematic corrections of order $m_\Lambda/M_{\Lambda_b}$ 
together with the perturbative corrections to the form-factor relations leads to a significant
effect in the ratio $H_L/H_T$ above $q^2=2$~{\rm GeV}$^2$, 
whereas the ratio $H_A/H_T$ is not very much affected.
In particular, we only find a small shift in the value of the forward-backward asymmetry zero,
\begin{align}
q_0^2 &= \left\{ \begin{array}{lr} 3.6~{\rm GeV}^2 & \qquad \mbox{\small (SCET limit),} \\
                                    3.4~{\rm GeV}^2 & \qquad\mbox{\small (incl.\ corrections).}
                  \end{array} \right.
\end{align}
Because of the small imaginary part of the term $(q^2 C_9^{\rm eff}(q^2) + 2 M_b M_{\Lambda_b} C_7^{\rm eff} )$
in the large-recoil region, the function $H_A/H_T$ also develops a pronounced minimum with $H_A \simeq -H_T$.
Again, its position is only slightly shifted from $q^2 \simeq 1.9~{\rm GeV}^2$ in the SCET limit, to $q^2 \simeq 1.7~{\rm GeV}^2$.
Notice that the function $\Delta \xi_\Lambda$, which describes the 
spectator corrections to the form factors, enters the above observables with
an additional suppression factor $2m_\Lambda/M_{\Lambda_b} \sim 40\%$, and therefore, even if -- as indicated --
we assign a large uncertainty to the ratio $\Delta \xi_\Lambda/\xi_\Lambda$, the considered ratios do not
change a lot.  The hard vertex corrections from the SCET matching coefficients $C_{f_i}$
and the purely kinematic corrections are thus responsible for the dominant numerical effect, together
with the unspecified uncertainties from non-factorizable and power corrections.

\clearpage

\section{Conclusions and Outlook}

In this article, we have systematically investigated the form factors
entering the baryonic $\Lambda_b \to \Lambda \ell^+ \ell^-$ transitions in the framework of
soft-collinear effective theory (SCET). As a starting point, we have introduced 
an improved form-factor parametrization, which leads to simple symmetry relations in
the limit of heavy $b$-quark mass $m_b$ and/or large recoil-energy $E_\Lambda$ to the $\Lambda$ baryon, and 
which yields simple expressions for partial decay widths and decay asymmetries.
We have shown that in the large recoil-energy limit, the 10 physical form factors for $\Lambda_b \to \Lambda$ 
transitions reduce to a single ``soft'' function $\xi_\Lambda(E_\Lambda)$, which can be defined as a 
matrix element of a universal decay current in SCET. The latter has been estimated from a sum-rule
analysis of an SCET correlation function, where the light $\Lambda$ baryon is 
interpolated by a suitable 3-quark current, and the heavy $\Lambda_b$ baryon is
described by its light-cone distribution amplitudes (LCDAs). We have studied the energy dependence
of the soft form factor, and performed a critical analysis of the uncertainties 
arising from the parameters used for the description of the hadronic continuum contribution to the sum rule, 
and for the model of the LCDAs. 
Compared to the recent measurement of the partially integrated $\Lambda_b \to \Lambda \mu^+\mu^-$
rate, we have found agreement within still \emph{large} experimental and theoretical uncertainties.

For phenomenological analyses related to precision tests of the SM or searches for
new physics, it is more convenient to study decay asymmetries, where -- to first approximation --
the dependence on hadronic form factors drops out in the large recoil-energy limit. In contrast
to the analogous mesonic transitions, both, the ratio $H_L/H_T$ of the longitudinal and transverse
decay rate, as well as the ratio $H_A/H_T$ defining the forward-backward asymmetry zero normalized
to the transverse rate, are independent of the hadronic form factors in the SCET limit.
A potentially important source of corrections 
arises from short-distance gluon exchange between the partonic $b \to s \ell^+\ell^-$ transition
and the spectator quarks in the baryons. We have shown that the leading effect can be described
by a hadronic matrix element of a particular sub-leading decay current in SCET. In contrast to
the mesonic $b \to s \ell^+\ell^-$ transitions, the so-defined
correction term $\Delta \xi_\Lambda$ cannot be calculated within the QCD-factorization approach,
because one of the two spectator quarks may still populate the kinematic endpoint region where
the resulting convolution integrals are ill-defined (in Appendix \ref{LambDAsimpl}, we briefly discuss
how this could be avoided by switching to a toy model with elementary light di-quark states in the
baryons).
Still, the function $\Delta \xi_\Lambda$ can be obtained from 
a sum-rule analysis of another SCET correlation function, and the contributions 
to the individual transition form factors can be identified. Numerically, we find that the corrections 
$\Delta \xi_\Lambda/\xi_\Lambda$ only amount to a few percent or less. The corresponding corrections
to the decay asymmetries have been estimated as well, including the effect of $\alpha_s$-corrections 
to the Wilson coefficients appearing in the matching of QCD decay currents onto the
leading SCET current, and kinematic corrections of order $m_\Lambda/M_{\Lambda_b}$.

Another source of (partially perturbatively calculable) corrections to $\Lambda_b\to\Lambda \ell^+\ell^-$
decay observables is related to so-called ``non-factorisable'' effects which cannot be
described in terms of $\Lambda_b \to \Lambda$ form factors. A systematic analysis of these
contributions -- following the analogous case of $B \to K^*\ell^+\ell^-$ decays in 
\cite{Beneke:2004dp} -- is left for future work. 
Finally, sub-leading terms in the SCET decay currents and SCET interaction terms between
soft and collinear fields will lead to power corrections involving sub-leading components
of the $\Lambda_b$ wave functions described by a number of new independent LCDAs. Since,
at the moment, only little is known about the partonic structure of the $\Lambda_b$ 
at sub-leading order, the non-perturbative power corrections remain an irreducible
source of hadronic uncertainties in rare exclusive $b$-quark decays.


\section*{Note Added}

The symmetry relations between baryonic form factors in the large-recoil limit
have also been discussed in a related recent paper in~\cite{Mannel:2011xx}. We thank
Thomas Mannel and Yu-Ming Wang for sharing their results with us
prior to publication. T.F.\ would also like to thank Yu-Ming~Wang for helpful discussions 
on the choice of interpolating currents.

\section*{Acknowledgements}

We thank Yu-Ming Wang for pointing out the inconsistencies in the derivation of the $\Lambda_b$ 
light-cone projector, which have been corrected in this updated version of the paper.
MWYY is supported by a Durham University Doctoral Fellowship.

\begin{appendices}

\section{Differential Decay Widths for $\Lambda_b \to \Lambda \mu^+\mu^-$}

\label{app:gammas}

In this appendix we provide the general formulas for
the differential decay widths for the radiative $\Lambda_b \to \Lambda \mu^+\mu^-$ transitions
in terms of the 10 helicity form factors defined in Sec.~\ref{sec:ff}. 
As usual, we consider the center-of-mass frame of the lepton-pair,
and define the angle $\theta$ between the $\Lambda_b$ baryon and
the \emph{positively} charged lepton. For simplicity, we consider massless
leptons, such that $q^2 = 2 \,  k_{\ell^+} \cdot k_{\ell^-}$. We then have
\begin{align}
 p_{\Lambda_b} \cdot k_{\ell^\pm} &= \frac{M_{\Lambda_b}^2 - m_\Lambda^2 + q^2 \mp \lambda \, \cos\theta}{4}
\,,
\qquad
 p_{\Lambda} \cdot k_{\ell^\pm} = \frac{M_{\Lambda_b}^2 - m_\Lambda^2 - q^2 \mp \lambda \, \cos\theta}{4}
 \,.
\end{align}
Here,
\begin{align}
\lambda \equiv \sqrt{s_+ s_-} &= \sqrt{((M_{\Lambda_b}+m_\Lambda)^2-q^2)\,(M_{\Lambda_b}-m_\Lambda)^2-q^2)} \,,
\end{align}
is the usual phase-space factor.
If we define
\begin{align}
 \frac{d^2\Gamma(\Lambda_b \to \Lambda \ell^+\ell^-)}{dq^2 \, d\cos\theta} 
& \equiv
\frac38 \left\{ (1+\cos^2\theta) \, H_T(q^2) + 2 \cos \theta \, H_A(q^2) + 2 (1-\cos^2\theta) \, H_L(q^2) \right\} \,,
\end{align}
and neglect non-factorisable contributions,
the different contributions to the differential decay rate can be written in terms of
the form factors in the helicity basis,
\begin{align}
 H_T(q^2) &= \frac{\lambda \, q^2 \, n}{96 \pi^3 M_{\Lambda_b}^3} \Bigg\{
s_- \left( 
\left| C_9^{\rm eff}(q^2) \, f_\perp 
+ \frac{2 M_b \, (M_{\Lambda_b}+ m_\Lambda) \, C_7^{\rm eff}}{q^2}  \, h_\perp \right|^2
+ \left| C_{10} \, f_\perp \right|^2 \right)
  \cr 
&  \phantom{\frac{\lambda \, q^2 \, n}{96 \pi^3 M_{\Lambda_b}^3}} \quad +
s_+ \left( 
\left| C_9^{\rm eff}(q^2) \, g_\perp 
+ \frac{2 M_b \, (M_{\Lambda_b}-m_\Lambda) \, C_7^{\rm eff}}{q^2}  \, \tilde h_\perp \right|^2
+ \left| C_{10} \, g_\perp \right|^2 \right)
\Bigg\} \,,
\\[0.3em]
 H_A(q^2) &= - \frac{\lambda^2 \,q^2 \, n }{48 \pi^3 M_{\Lambda_b}^3} \,  {\rm Re}
\left[ \left( C_9^{\rm eff}(q^2) \, f_\perp 
+ \frac{2 M_b \, (M_{\Lambda_b}+ m_\Lambda) \, C_7^{\rm eff}}{q^2}  \, h_\perp \right)^*
\left( C_{10} g_\perp \right)
\right.
\cr 
& \phantom{\frac{\lambda^2 \, n \, q^2 }{48 \pi^3 M_{\Lambda_b}^3} \,  {\rm Re}} \qquad
 \left.
+\left( C_9^{\rm eff}(q^2) \, g_\perp 
+ \frac{2 M_b \, (M_{\Lambda_b}- m_\Lambda) \, C_7^{\rm eff}}{q^2}  \, \tilde h_\perp \right)^*
\left( C_{10} f_\perp \right)
\right] \,,
\\[0.3em] 
 H_L(q^2) &= \frac{\lambda \, n}{192 \pi^3 M_{\Lambda_b}^3} \Bigg\{
s_- \, (M_{\Lambda_b}+ m_\Lambda)^2 \left( 
\left|C_9^{\rm eff}(q^2) \, f_+ 
+ \frac{2 M_b\, C_7^{\rm eff}}{M_{\Lambda_b}+ m_\Lambda}  \, h_+ \right|^2
+ \left| C_{10} \, f_+ \right|^2\right)
 \cr 
&   \phantom{\frac{\lambda \, n}{192 \pi^3 M_{\Lambda_b}^3}} \quad +
s_+ \, (M_{\Lambda_b}- m_\Lambda)^2 \left(
\left| C_9^{\rm eff}(q^2) \,  g_+ 
+ \frac{2 M_b \, C_7^{\rm eff}}{M_{\Lambda_b}-m_\Lambda}  \, \tilde h_+ \right|^2
+ \left| C_{10} \, g_+ \right|^2 \right) \Bigg\} \,.
\cr &
\label{eqs:obsall}
\end{align}
where
\begin{align}
 n &= \frac{\alpha^2 \, G_F^2}{8 \pi^2} \, |V_{ts} V_{tb}|^2 \,.
\end{align}
The functions become particularly simple in the SCET limit, where
\begin{align}
 H_T(q^2) &\simeq \frac{\lambda^2 \, q^2 \, n}{48 \pi^3 M_{\Lambda_b}^3} \, |\xi_\Lambda(n_+p')|^2
\Bigg\{
\left| C_9^{\rm eff}(q^2)  
+ \frac{2 M_b M_{\Lambda_b} \, C_7^{\rm eff}}{q^2}  \right|^2
+ \left| C_{10} \right|^2 \Bigg\} \,,
\\[0.3em]
 H_A(q^2) &\simeq - \frac{\lambda^2 \, q^2 \, n }{24 \pi^3 M_{\Lambda_b}^3} \, |\xi_\Lambda(n_+p')|^2 \,  {\rm Re}
\left[ \left( C_9^{\rm eff}(q^2) 
+ \frac{2 M_b M_{\Lambda_b} \, C_7^{\rm eff}}{q^2}  \right)^*
 C_{10} \right]
 \,,
\\[0.3em] 
 H_L(q^2) &\simeq \frac{\lambda^2 \, n}{96 \pi^3 M_{\Lambda_b}}  \, |\xi_\Lambda(n_+p')|^2 \Bigg\{
\left|C_9^{\rm eff}(q^2) 
+ \frac{2M_b}{M_{\Lambda_b}} \, C_7^{\rm eff} \right|^2
+ \left| C_{10} \right|^2
 \Bigg\} \,.
\end{align}

\section{Alternative Form-Factor Parametrizations}

\subsection{Convention by Chen and Geng}

\label{app:otherff}

The form factors in \cite{Chen:2001zc}, which have been  commonly used
in the recent literature, are related to ours as follows.
For the vector form factors, we obtain
\begin{align}
 f_0 &= f_1 + \frac{q^2}{M_{\Lambda_b}-m_\Lambda} \, f_3 \,,
\qquad
 f_+ = f_1 - \frac{q^2}{M_{\Lambda_b}+m_\Lambda} \, f_2  \,,
\qquad
 f_\perp = f_1 - (M_{\Lambda_b}+m_\Lambda) \, f_2
\,.
\end{align}
Similarly, for the axial-vector form factors, one gets
\begin{align}
 g_0 &= g_1 - \frac{q^2}{M_{\Lambda_b}+m_\Lambda} \, g_3 \,,
\qquad
 g_+ = g_1 + \frac{q^2}{M_{\Lambda_b}-m_\Lambda} \, g_2  \,,
\qquad
 g_\perp = g_1 + (M_{\Lambda_b}-m_\Lambda) \, g_2
\,.
\end{align}
The tensor and pseudo-tensor form factors are related by
\begin{align}
 h_+ = f_2^T - \frac{M_{\Lambda_b}+m_\Lambda}{q^2} \, f_1^T  \,,
\qquad
 h_\perp = f_2^T - \frac{1}{M_{\Lambda_b}+m_\Lambda} \, f_1^T
\,,
\end{align}
and
\begin{align}
 \tilde h_+ = g_2^T + \frac{M_{\Lambda_b}-m_\Lambda}{q^2} \, g_1^T  \,,
\qquad
 \tilde h_\perp = g_2^T + \frac{1}{M_{\Lambda_b}-m_\Lambda} \, g_1^T
\,.
\end{align}

\subsection{Symmetry-Based Form-Factor Parametrization}

\label{app:altpar}

An alternative parametrization considers the different projections of the decay current
in the heavy-quark and/or large-energy limit, respectively.
On the heavy-quark side, we consider the heavy-baryon velocity $v^\mu = p^\mu/M_{\Lambda_b}$ such that 
$\slash v \, u_{\Lambda_b}(p) = u_{\Lambda_b}(p)$. 
Also taking into account the projections on the light-quark side (using parity invariance of strong interactions), 
we end up with the general expression
\begin{align}
 \langle \Lambda(p',s')| \bar q \, \Gamma \,  b|\Lambda_b(p,s)\rangle &= 
\xi_{ij}^{(\pm)}(v,p') \, \bar u_\Lambda(p',s') \left\{ \Gamma_i \, \frac{\slash n_\pm \slash n_\mp}{4} 
\, \Gamma  \, \Gamma_j \right\}  u_{\Lambda_b}(p,s)
\label{ansatz}
 \end{align}
where the basis of Dirac matrices can be chosen as 
\begin{align}
 \Gamma_i &= \{ 1 , \gamma_5, \gamma_\perp^\alpha \} \,, \qquad 
 \Gamma_j = \{ 1 , \gamma_5, \vec \gamma_\mu, \vec\gamma_\mu\gamma_5 \} \,,
\end{align}
and $\gamma_\perp^\alpha=\gamma^\alpha - \frac{\slash n_+}{2} n_-^\alpha - \frac{\slash n_-}{2} n_+^\alpha $, while
$\vec \gamma_\mu = \gamma_\mu - \slash v \, v_\mu$ etc. Here and in the following, 
we consider a frame where $\slash p'_\perp=0$ and $v^\mu = (n_-^\mu + n_+^\mu)/2$.
The non-vanishing form factors  are
\begin{align}
 \xi_{11}^{(\pm)}(v , p') & \equiv A^{(\pm)}( v \cdot p') \sim {\cal O}(1) \,,
\qquad
 \xi_{13}^{(\pm)}(v , p')  \equiv \frac{p'{}^\mu}{v \cdot p'} \, B^{(\pm)}( v \cdot p')  \sim {\cal O}(\epsilon) \,,
\cr 
 \xi_{22}^{(\pm)}(v , p') & \equiv C^{(\pm)}( v \cdot p') \sim {\cal O}(\epsilon) \,,
\qquad
 \xi_{24}^{(\pm)}(v , p')  \equiv \frac{p'{}^\mu}{v \cdot p'} \, D^{(\pm)}( v \cdot p')  \sim {\cal O}(\epsilon) \,,
\cr 
\xi_{33}^{(\pm)}(v , p') & \equiv \delta_\alpha^\mu \, E^{(\pm)}( v \cdot p') \sim {\cal O}(\epsilon) \,,
\cr 
\xi_{34}^{(\pm)}(v , p')  &\equiv i\epsilon_\alpha{}^{\mu\rho\sigma} \, \frac{v_\rho \, p'_\sigma }{v \cdot p'} \, F^{(\pm)}( v \cdot p') \sim {\cal O}(\epsilon) \,,
\qquad
\end{align}
From the above 12 form factors, again, only 10 are independent, after the e.o.m.\ constraints
have been taken into account. 
Here, the indicated suppression of the form factors with $\epsilon=\Lambda/M$
refers to the violation of the heavy-quark spin symmetry.
In addition, in the large recoil limit the contributions from the 
form factors with an index ``$-$'' are additionally suppressed.
Therefore, we may neglect the 5 form factors $B^{(-)}$ through $F^{(-)}$, which is a good approximation, because
\begin{itemize}
 \item In the HQET limit, $v \cdot p' \sim {\cal O}(m_\Lambda)$, their contribution is suppressed
       at least by a factor $\Lambda/M$.
 \item In the SCET limit, $n_+  p' \sim {\cal O}(M_{\Lambda_b})$, their contribution is suppressed by
       at least a factor $(\Lambda/M)^2$ (for non-factorizable effects) or $\alpha_s$ (for factorizable effects, see below). 
\end{itemize}
We thus end up with a rather efficient description which combines the symmetry constraints in both
cases and allows one to systematically take into account sub-leading corrections
in the large-recoil limit, which are partially calculable in the framework of QCD factorization
or light-cone sum rules.
In this approximation, the 10 physical helicity form factors are related by 5 equations (for vanishing light quark masses, $m_s \to 0$),
\begin{align}
f_0 &= 
\frac{M_{\Lambda_b}+m_\Lambda}{M_{\Lambda_b}-m_\Lambda}
\, \frac{n_+p'-m_\Lambda}{n_+p'+m_\Lambda} \, f_+
+\frac{M_{\Lambda_b}-n_+p'}{M_{\Lambda_b}-m_\Lambda} 
 \left( g_\perp - \frac{n_+p'-m_\Lambda}{n_+p'+m_\Lambda} \, f_\perp \right) 
\,,
\cr
 g_0 &= \frac{M_{\Lambda_b}-m_\Lambda}{M_{\Lambda_b}+m_\Lambda}
\, \frac{n_+p'+m_\Lambda}{n_+p'-m_\Lambda} \, g_+ 
+ \frac{M_{\Lambda_b}-n_+p'}{M_{\Lambda_b}+m_\Lambda} 
 \left( f_\perp- \frac{n_+p'+m_\Lambda}{n_+p'-m_\Lambda} \, g_\perp \right)
\,,
\cr 
 \tilde h_\perp &= 
\frac{M_{\Lambda_b}+m_\Lambda}{M_{\Lambda_b}-m_\Lambda}
\, \frac{n_+p'-m_\Lambda}{n_+p'+m_\Lambda} \, h_\perp
+\frac{M_{\Lambda_b}-n_+p'}{M_{\Lambda_b}-m_\Lambda} 
 \left( g_\perp - \frac{n_+p'-m_\Lambda}{n_+p'+m_\Lambda} \, f_\perp \right) \,, 
\end{align}
and
\begin{align}
 h_+ &= 
\frac{M_{\Lambda_b}+m_\Lambda}{M_b} \, f_+ 
+ \frac{n_+p'-\Lambda}{M_b}
 \left( f_\perp - \frac{n_+p'+m_\Lambda}{n_+p'-m_\Lambda}  \, g_\perp
\right)
\,,
\cr 
\tilde h_+ &=
\frac{M_{\Lambda_b}-m_\Lambda}{M_b} \, g_+ 
+ \frac{n_+p'-\Lambda}{M_b}
 \left( g_\perp - \frac{n_+p'-m_\Lambda}{n_+p'+m_\Lambda}  \, f_\perp
\right)
 \,.
\end{align}

\section{Corrections to SCET Symmetry Relations}

\label{app:corr}

\subsection{Hard Vertex Corrections}

The hard vertex corrections to the individual QCD decay currents have
been discussed before \cite{Beneke:2000wa,Bauer:2000yr}. From the general
1-loop result in Eq.~(28) in \cite{Beneke:2000wa} we can deduce the
corrections to the individual form factors in the helicity basis, 
$f_i = C_{f_i} \, \xi_\Lambda +\ldots$. Defining the renormalization 
scheme through $C_{f_+} = C_{g_+}  \equiv 1$, this leads to
\begin{align}
& C_{f_0} = C_{g_0}  = 1 + \frac{\alpha_s C_F}{4\pi} \, 2  (1-L) \,, \qquad
 C_{f_\perp}=C_{g_\perp}  = 1 + \frac{\alpha_s C_F}{4\pi} \, L \,, 
\end{align}
and
\begin{align}
& C_{h_+} = C_{\tilde h_+}  = 1 + \frac{\alpha_s C_F}{4\pi} \left( \ln \frac{M_b^2}{\mu^2} -2 (1-L) \right) \,, \qquad
 C_{h_\perp} = C_{\tilde h_\perp}  = 1 + \frac{\alpha_s C_F}{4\pi} \left( \ln \frac{M_b^2}{\mu^2} -2  \right) \,,
\end{align}
with the abbreviation
$$
L \equiv - \frac{M_b^2-q^2}{q^2} \, \ln \left( 1- \frac{q^2}{M_b^2} \right) \,.
$$

\subsection{Hard-Collinear Gluon Exchange}

We consider the tree-level matching (in light-cone gauge), following \cite{Beneke:2000wa}
\begin{align}
 \bar q \, \Gamma \, Q_v & \simeq\, \bar \xi \, \tilde \Gamma \, h_v 
  - \frac{1}{n_+p'} \, \bar \xi \, g \slash A_\perp \, \frac{\slash n_+}{2} \, \Gamma\, h_v
 - \frac{1}{M_b} \, \bar \xi \, \Gamma \, \frac{\slash n_-}{2} \, g \slash A_\perp \, h_v + \ldots
\end{align}
The hard-scattering contributions to the individual form factors in the large-recoil limit defined above can then
 be identified by means of (\ref{deltasoft}) and 
setting $m_\Lambda \to 0$ and $M_{\Lambda_b} \to M_b \equiv M$. This is equivalent to using
\begin{align}
 A^{(-)} &\simeq - \frac{2 M}{m_\Lambda} \, \Delta \xi_\Lambda  \,,  \qquad
 E^{(+)} = F^{(+)} = \frac12 \, \Delta \xi_\Lambda
\end{align}
in (\ref{ansatz}).
For the scalar and vector form factors, this yields
\begin{align}
 f_0(q^2) & \simeq 
 C_{f_0} \, \xi_\Lambda(n_+p') - \frac{2M}{n_+p'} \, \Delta \xi_\Lambda(n_+p') \,,
\cr 
 f_+(q^2) &\simeq C_{f_+} \, \xi_\Lambda(n_+p') - 2 \left(2 - \frac{M}{n_+p'} \right) \, \Delta \xi_\Lambda(n_+p') \,,
\cr 
f_\perp(q^2) & \simeq 
 C_{f_\perp} \, \xi_\Lambda(n_+p') + \frac{2M}{n_+p'} \, \Delta \xi_\Lambda(n_+p') \,,
\end{align}
where $C_i=C_i(\mu,n_+p')$ denote the hard vertex coefficients as derived above. 
Similar relations can be obtained for the axial-vector and tensor form factors,
\begin{align}
 g_0(q^2) & \simeq 
 C_{g_0} \, \xi_\Lambda(n_+p') + \frac{2M}{n_+p'} \, \Delta \xi_\Lambda(n_+p') \,,
\cr 
 g_+(q^2) &\simeq C_{g_+} \, \xi_\Lambda(n_+p') + 2 \left(2 - \frac{M}{n_+p'} \right) \, \Delta \xi_\Lambda(n_+p') \,,
\cr 
g_\perp(q^2) & \simeq 
 C_{g_\perp} \, \xi_\Lambda(n_+p') - \frac{2M}{n_+p'} \, \Delta \xi_\Lambda(n_+p') \,,
\end{align}
and
\begin{align}
 h_+(q^2) &\simeq C_{h_+} \, \xi_\Lambda(n_+p') + \frac{2M}{n_+p'} \, \Delta \xi_\Lambda(n_+p') \,,
\cr 
h_\perp(q^2) & \simeq 
 C_{h_\perp} \, \xi_\Lambda(n_+p') - 2 \left(1- \frac{M}{n_+p'} \right)  \Delta \xi_\Lambda(n_+p') \,,
\end{align}
and
\begin{align}
 \tilde h_+(q^2) &\simeq C_{\tilde h_+} \, \xi_\Lambda(n_+p') - \frac{2M}{n_+p'} \, \Delta \xi_\Lambda(n_+p') \,,
\cr 
 \tilde h_\perp(q^2) & \simeq 
 C_{\tilde h_\perp} \, \xi_\Lambda(n_+p') +  2 \left(1- \frac{M}{n_+p'} \right) \Delta \xi_\Lambda(n_+p') \,.
\end{align}

\subsection{Form-Factor Relations to ${\cal O}(\alpha_s)$ Accuracy}

To first order in the strong coupling constant, the hard vertex corrections and the spectator scattering
corrections only provide 5 independent Dirac structures. As a consequence, after inclusion of ${\cal O}(\alpha_s)$
corrections, from the 10 helicity form factors only 5 are still linearly independent.\footnote{A similar effect was 
observed for $B \to V=\rho,K^* \ldots$ transitions, where among the 7 physical form factors 2 symmetry relations
remain at ${\cal O}(\alpha_s)$ \cite{Beneke:2000wa}. 
Symmetry arguments based on the helicity conservation of the light quark in
short-distance interactions can be found in \cite{Burdman:2000ku}.
For $B$-meson decays into light pseudoscalars no such relations remain, because
there are only 3 physical form factors to start with in the first place.} The 5 symmetry relations which
are unaffected by ${\cal O}(\alpha_s)$ radiative corrections can be summarized as
\begin{align}
& \frac{f_0 + h_+}{g_0 + \tilde h_+} 
= \frac{f_\perp - h_+}{g_\perp - \tilde h_+} 
= \frac{f_+ + h_+ - 2 h_\perp}{g_+ + \tilde h_+ -2 \tilde h_\perp}
\cr 
& = - \frac{2 (f_+-f_\perp)+ h_+ - h_\perp}{2 (g_+-g_\perp)+ \tilde h_+ - \tilde h_\perp}
= - \frac{M^2-2 q^2}{M^2} \, \frac{h_+ - \tilde h_+}{f_+-g_+}
\ = \ 1 \,.
\end{align}

\clearpage

\section{Light-Cone Distribution Amplitudes}

\label{app:DAs}

Light-cone distribution amplitudes (LCDAs) are introduced as matrix
elements of non-local QCD light-ray operators between the considered
baryon states and the vacuum.

\subsection{Distribution Amplitudes for the $\Lambda_b$ baryon}

\label{app:DA1}

For the heavy $\Lambda_b$ baryon, we follow the definitions in \cite{Ball:2008fw}
and consider the following two projections (two others are not shown),
\begin{align}
\label{def:DA}
 \epsilon^{abc} \, \langle 0| \left(u^a(t_1 n_-)\, C \,\gamma_5 \slash n_-  \, d^b(t_2 n_-)\right) h^c_{v}(0)
  |\Lambda_b(v,s)\rangle &= f^{(2)}_{\Lambda_b} \, \Psi_2(t_1,t_2) \, u_{\Lambda_b}(v,s)\,,
\nonumber\\ 
\epsilon^{abc} \, \langle 0| \left(u^a(t_1 n_-)\, C \,\gamma_5 \slash n_+  \, d^b(t_2 n_-)\right) h^c_{v}(0)
  |\Lambda_b(v,s)\rangle &= f^{(2)}_{\Lambda_b} \, \Psi_4(t_1,t_2) \, u_{\Lambda_b}(v,s)\,.
\end{align}
The so-defined LCDAs in position space have a Fourier expansion, 
\begin{align}
 \Psi_i(t_1,t_2) & = \int_0^\infty d\omega_1 \int_0^\infty d\omega_2 \, e^{-i (t_1 \omega_1+t_2 \omega_2)} \,\psi_i(\omega_1,\omega_2)
\cr 
 &= \int_0^\infty d\omega \, \omega \int_0^1 du \, e^{-i \omega (t_1 u+t_2 \bar u)} \,\tilde \psi_i(\omega,u) \,.
\end{align}
Here, the first alternative refers to a function of the two light-cone momenta $\omega_{1,2}=(n_-k_{1,2})$ 
of the two light quarks in the heavy baryon, while the second alternative considers the total light-cone
momentum $\omega=\omega_1+\omega_2$ and the momentum \emph{fractions} $u=\omega_1/\omega$, 
$\bar u=1-u=\omega_2/\omega$ (notice the additional factor of $\omega$ in the Fourier integral
in the latter case). The normalization factors $f_{\Lambda_b}^{(1,2)}$ have mass-dimension~3 
and are scale-dependent. For numerical estimates, we will use 
$f_{\Lambda_b}^{(i)} \simeq 0.030 \pm 0.005~{\rm GeV}^3$. The LCDAs $\psi_i(\omega_1,\omega_2)$ 
in momentum space have mass-\mbox{dimension $(-2)$} and are scale-dependent, too. More details can be
found in \cite{Ball:2008fw}.

The above definitions can be converted into momentum-space
representations for the $\Lambda_b$ distribution amplitudes,
following the analogous procedure that has been explained in detail for
the $B$-meson LCDA in \cite{Beneke:2000wa}.
Taking an arbitrary light-like vector $y^\mu$ and defining $t=v \cdot y
$, we can write the most general Lorentz decomposition in the heavy-quark
limit,
\begin{align} \label{def:DA2}
& \epsilon^{abc} \, \langle 0| \left(u^a (\tau_1 y) \, C \gamma_5
\gamma_\mu \, d^b(\tau_2 y)\right) h^c_{v}(0)
  |\Lambda_b(v,s)\rangle 
\cr 
=& f^{(2)}_{\Lambda_b} \left( v_\mu \, \Psi_2(\tau_1,\tau_2)  +
\frac{\Psi_4(\tau_1,\tau_2)- \Psi_2(\tau_1,\tau_2)}{2t} \, y_\mu \right)
u_{\Lambda_b}(v,s)\,.
\end{align}
This can be turned into 
\begin{align} 
& \epsilon^{abc} \, \langle 0| \left(u_\alpha^a (\tau_1 y) \, d^b_
\beta(\tau_2 y)\right) h^c_{v}(0)
  |\Lambda_b(v,s)\rangle 
\cr 
=& \frac{f^{(2)}_{\Lambda_b}}{4} u_{\Lambda_b}(v,s) \,
\left[ \left( \slash v \, \Psi_2(\tau_1,\tau_2)  + \frac{\Psi_4(\tau_1,
\tau_2)- \Psi_2(\tau_1,\tau_2)}{2t} \, \slash y \right)  \gamma_5 C^{-1}
\right]_{\beta\alpha} 
\cr 
& \quad + \mbox{2 more terms.} \label{eq:projector}
\end{align}

In the convolution with hard-scattering kernels that have a
power expansion in the transverse momenta $k_{1\perp}$ and $k_{2\perp}$ of
the two light quarks in the $\Lambda_b$ baryon, and which have
a corresponding sub-sub-leading dependence on $(n_+k_i)$, 
the most general momentum-space projector 
\begin{align}
  \frac{f^{(2)}_{\Lambda_b}}{4} u_{\Lambda_b}(v,s) \, \tilde M(k_1,k_2)_{\beta\alpha} \Big|_{k_i = \omega_i n_+/2}
\end{align}
reads \cite{Feldmann:2012xx}
 \begin{align} 
\tilde M(k_1,k_2) = & 
 \left( \frac{\slash n_+}{2} \, \psi_2(\omega_1,\omega_2) +
\frac{\slash n_-}{2} \, \psi_4(\omega_1,\omega_2)\right.
\cr 
& \qquad \quad \left. 
 - \frac{1}{2} \, \gamma_\mu^\perp \, \int_0^{\omega_1} d\eta_1 \left( \psi_{42}^{(1)}(\eta_1,\omega_2) - \psi_X(\eta_1,\omega_2) \right)
\frac{\slash n_+\slash n_-}{4}
\, \frac{\partial}{\partial k_{1\mu}^\perp} 
\right. \cr & \qquad \quad \left.   - \frac{1}{2} \, \gamma_\mu^\perp \, \int_0^{\omega_1} d\eta_1 \left( \psi_{42}^{(1)}(\eta_1,\omega_2) + \psi_X(\eta_1,\omega_2) \right)
\frac{\slash n_-\slash n_+}{4}
\, \frac{\partial}{\partial k_{1\mu}^\perp} 
\right. \cr & \qquad \quad \left.  - \frac{1}{2} \, \gamma_\mu^\perp \, \int_0^{\omega_2} d\eta_2 \left( \psi_{42}^{(2)}(\omega_1,\eta_2) - \psi_X(\omega_1,\eta_2) \right)
\frac{\slash n_-\slash n_+}{4}
\, \frac{\partial}{\partial k_{2\mu}^\perp} 
\right. \cr & \qquad \quad \left.    - \frac{1}{2} \, \gamma_\mu^\perp \, \int_0^{\omega_2} d\eta_2 \left( \psi_{42}^{(2)}(\omega_1,\eta_2) + \psi_X(\omega_1,\eta_2) \right)
\frac{\slash n_+\slash n_-}{4}
\, \frac{\partial}{\partial k_{2\mu}^\perp} 
\right)
\gamma_5 C^{-1}
\cr & \qquad
 + \mbox{2 more terms.} 
\end{align}
Here, $\psi_{42}^{(2)}(\omega_1,\omega_2) = \psi_{42}^{(1)}(\omega_2,\omega_1)$ and
$\psi_X(\omega_1,\omega_2) =\psi_X(\omega_2,\omega_1)$ and
\begin{align}
 \psi_{42}^{(1)}(\omega_1,\omega_2) + \psi_{42}^{(2)}(\omega_1,\omega_2) &= \psi_4(\omega_1,\omega_2)-\psi_2(\omega_1,\omega_2) \,.
\end{align}

From this we see that $\psi_{2,4}$ play the analogous role as
$\phi_\pm^B$ for the $B$-meson. 
%
The asymmetric combination of $\psi_{42}^{(1)}$ and $\psi_{42}^{(2)}$, as well as $\psi_X$
do not contribute in the collinear limit (\ref{eq:projector}). 
However, they do contribute to the correlator used for the sum-rule estimate of $\Delta \xi_\Lambda$.
They also allow one to derive approximate Wandzura-Wilczek relations from the equations of motion,
\begin{align}
 \slash k_2  \tilde M(k_1,k_2) \simeq \tilde M(k_1,k_2) \slash k_1 \simeq 0 \,,
\end{align}
\noindent in the limit of vanishing LCDAs with $n>3$ partons.

Parametrizations for the functional form of the LCDAs have been derived from a sum-rule analysis in \cite{Ball:2008fw}.
In this paper, we will use a simple model which is based on an exponential ansatz suppressing 
large values of $(v \cdot k_{1,2})$, where $k_{1,2}$ are the on-shell momenta for the light quarks 
in the $\Lambda_b$,
\begin{align}
  f(v \cdot k_1, v\cdot k_2) &:= \frac{1}{\omega_0^6} \, e^{- (v \cdot k_1 + v \cdot k_2)/\omega_0} \,.
\end{align}
where $\omega_0 \sim \Lambda_{\rm had}$ is a measure for the typical momentum of the di-quark.
We then may use 
\begin{align}
 \psi_2(\omega_1,\omega_2) &= \int_0^\infty dk_{1\perp}^2 \, dk_{2\perp}^2 \, 
  f\left(\omega_1 + \frac{k_{1\perp}^2}{\omega_1}, \omega_2 + \frac{k_{2\perp}^2}{\omega_2}\right)
 = \frac{\omega_1 \,\omega_2\, e^{-(\omega_1+\omega_2)/\omega_0}}{\omega_0^4} \,,
\\[0.2em]
 \psi_4(\omega_1,\omega_2) &= \int_0^\infty dk_{1\perp}^2 \, dk_{2\perp}^2 \, \frac{k_{1\perp}^2}{\omega_1^2} \, \frac{k_{2\perp}^2}{\omega_2^2} \,
  f\left(\omega_1 + \frac{k_{1\perp}^2}{\omega_1}, \omega_2 + \frac{k_{2\perp}^2}{\omega_2}\right)
 = \frac{e^{-(\omega_1+\omega_2)/\omega_0}}{\omega_0^2} \,,
\label{Lamb:LCDA}
\end{align}
where the pre-factors in the integrand of the second line correspond to the ratios $(n_+k_i)/(n_-k_i)$ taking into account
that $\psi_2$ and $\psi_4$ change their role when switching $n_+ \leftrightarrow n_-$.
Also (see \cite{Feldmann:2012xx} for details),
\begin{align}
 \psi_{42}^{(1)}(\omega_1,\omega_2) &= \frac{\partial}{\partial \omega_1} \int_0^\infty dk_{1\perp}^2 \, dk_{2\perp}^2 \, \frac{k_{1\perp}^2}{2\omega_1} \left(1+\frac{k_{2\perp}^2}{\omega_2^2} \right) \,
  f\left(\omega_1 + \frac{k_{1\perp}^2}{\omega_1}, \omega_2 + \frac{k_{2\perp}^2}{\omega_2}\right)
 \cr
 &= \frac{(\omega_0-\omega_1)(\omega_0+\omega_2)}{2\omega_0^4} \, e^{-(\omega_1+\omega_2)/\omega_0} \,,
\\
 \psi_X(\omega_1,\omega_2)          &= -\frac{\partial}{\partial \omega_1} \int_0^\infty dk_{1\perp}^2 \, dk_{2\perp}^2 \, \frac{k_{1\perp}^2}{2\omega_1} \left(1-\frac{k_{2\perp}^2}{\omega_2^2} \right) \,
  f\left(\omega_1 + \frac{k_{1\perp}^2}{\omega_1}, \omega_2 + \frac{k_{2\perp}^2}{\omega_2}\right)
 \cr
 &= \frac{(\omega_0-\omega_1)(\omega_0-\omega_2)}{2\omega_0^4} \, e^{-(\omega_1+\omega_2)/\omega_0} \,.
\end{align}
For later use, we also introduce the abbreviations
\begin{align}
 G(\omega_1,\omega_2) &= \int_0^{\omega_1} d\eta_1
 \left( \psi_{42}^{(1)}(\eta_1,\omega_2) - \psi_X(\eta_1,\omega_2) \right) \to
  \frac{\omega_1\omega_2}{\omega_0^3} \, e^{-(\omega_1+\omega_2)/\omega_0} \,,
\\
 H(\omega_1,\omega_2) &= \int_0^{\omega_2} d\eta_2 \left( \psi_{42}^{(2)}(\omega_1,\eta_2) + \psi_X(\omega_1,\eta_2) \right) \to
  \frac{\omega_2}{ \omega_0^2} \, e^{-(\omega_1+\omega_2)/\omega_0} \,.
\end{align}

The parameter $\omega_0$ has been estimated in \cite{Ball:2008fw} from a sum-rule analysis of $\tilde\psi_2(\omega,u)$
(also including corrections from higher-order Gegenbauer polynomials as a function of $(2u-1)$),
and rather small values of order $200$~MeV or so have been found. In our numerical analysis in the main body of the text,
we will use a somewhat higher value ($300$~MeV) as our default, but consider a rather large uncertainty associated to it.

\subsection{Simplified Set-Up with Scalar Di-quark}

\label{LambDAsimpl}

For a simplified picture, one may also approximate the dynamics of the two light quarks in the $\Lambda_b$ baryon
by an elementary 
scalar di-quark field $\varphi^a(x)$ in the $\bar 3$ representation of $SU(3)_C$. 
In the HQET limit, the $\Lambda_b$ baryon could then be described
by a single LCDA, defined as ($t=v\cdot z$)
\begin{align}
 \label{def:DAsimple}
  \langle 0| \varphi^a(z) \,  h^a_{v}(0)
  |\Lambda_b(v,s)\rangle &= \hat f_{\Lambda_b} \, \Psi_{\Lambda_b}(t) \, u_{\Lambda_b}(v,s)\,,
\end{align}
and
\begin{align}
\Psi_{\Lambda_b}(t) & = \int_0^\infty d\omega \, e^{-i t \omega} \,\phi_{\Lambda_b}(\omega) \,.
\end{align}
Here $\hat f_{\Lambda_b}$ has mass-dimension $+1$, and $\psi_{\Lambda_b}(\omega)$ has mass-dimension
$-1$. The momentum-space projector in this case simply reads
\begin{align}
 \hat f_{\Lambda_b} \,\phi_{\Lambda_b}(\omega) \, u_{\Lambda_b}(v,s) \,.
\end{align}

Similarly, the $\Lambda$ baryon can be approximately described 
by two LCDAs, defined as 
\begin{align}
 \label{def:DAsimple2}
  \langle \Lambda(p',s')| \bar s^a(z) \varphi^a{}^\dagger (0)  |0\rangle 
 &= \int_0^1 du \, e^{i u \, p' \cdot z} \,
    \bar u_{\Lambda}(p',s') \left( \hat f_\Lambda^{(1)} \, \phi_\Lambda^{(1)}(u) 
 - \sigma_{\mu\nu} p'{}^\mu z^\nu \,  \hat f_\Lambda^{(2)} \, \phi_\Lambda^{(2)}(u)
 \right)
\end{align}
which corresponds to a momentum-space projector
\begin{align}
 \bar u_{\Lambda}(p',s') \left( \hat f_\Lambda^{(1)} \, \phi_\Lambda^{(1)}(u) 
  - \frac{i}{2} \, \sigma_{\mu\nu} \hat f_\Lambda^{(2)} \left\{
n_-^\mu n_+^\nu \,  \phi_\Lambda^{(2)}{}'(u)
 - p'{}^\mu \, \frac{\partial}{\partial k_\perp{}_\nu} \,  \phi_\Lambda^{(2)}{}(u) \right\}\right) \,.
\end{align}

\paragraph{Soft form factor from simplified set-up:}

We may use the di-quark approximation as a toy model, to obtain alternative expressions for
the transition form factors from SCET sum rules. To this end,  we consider a correlation function involving the
interpolating current
\begin{align}
 \hat J_\Lambda(x) & = \varphi^a(x) \, s^a(x)
\label{Jhat}
\end{align}
with 
\begin{align}
 \langle 0| \hat J_\Lambda(0)| \Lambda(p',s')\rangle &= \hat f_\Lambda \,  u_\Lambda(p',s') 
\,.
\end{align}
The remaining calculation is analogous to the realistic case considered in Sec~\ref{sec:soft}, and yields the  LO sum rule
\begin{align}
 e^{-m_\Lambda^2/(n_+p') \hat \omega_M}  \, \hat f_\Lambda \, \xi_\Lambda(n_+p') 
&= \hat f_{\Lambda_b} \int_0^{\hat\omega_s} d\omega \, \phi_{\Lambda_b}(\omega) \, e^{-\omega/\hat\omega_M} \,,
\end{align}
with an according new threshold parameter $\hat\omega_s$ and Borel parameter $\hat\omega_M$.

\paragraph{Hard-collinear gluon correction from simplified set-up:}

In the simplified toy model, as before, we define the correlation function using the
interpolating current in (\ref{Jhat}),
\begin{align}
 \hat \Pi_\Lambda^\mu(n_-p') & \equiv i \int d^4x \, e^{ip'x} \langle 0| T \left[ \frac{\slash n_+ \slash n_-}{4} \, \hat J_\Lambda(x) \left[ \bar s(0) \, \tilde \Gamma 
  \, g A^\mu_\perp(0) \, b(0) \right] \right] |\Lambda_b(p)\rangle \,.
\end{align}
Evaluating the Feynman diagram (using scalar QCD for the di-quark in the $\bar 3$ representation), we obtain 
\begin{align}
 \hat \Pi_\Lambda^\mu (n_-p') & = - i g_s^2 C_F \, \hat f_{\Lambda_{b}} \, \int_0^\infty \!d\omega \,\phi_{\Lambda_b}(\omega)
\int \!\frac{d^D l}{(2\pi)^D} \,
\frac{1}{\left[l_\perp^2+ (n_+l)(n_-l)\right]} \, \slash l_\perp  l_\perp^\mu \, 
 \tilde\Gamma \, u_{\Lambda_{b}}(v,s)
\cr
 & \qquad \times \, \frac{1}{\left[l_\perp^2 + (n_+l)(n_-l -  \omega)\right]} \, 
    \frac{1}{\left[l_\perp^2 + (n_+l+n_+p')(n_-l+n_-p'-\omega)\right]}  \,.
\end{align}
The correlator can be calculated as before, leading to
\begin{align}
 \hat \Pi_\Lambda^\mu (n_-p') & = 
-  g_s^2 C_F \, \hat f_{\Lambda_{b}}\, \gamma^\mu_\perp \, \tilde \Gamma  \,u_{\Lambda_{b}}(v,s) \,
\int_0^\infty \!d\omega  \, \phi_{\Lambda_b}(\omega) \, \frac1{2} \, \int_0^1 dz
\cr
 & \quad \times  \int\! \frac{d^{D-2}l_\perp}{(2\pi)^{D-1}}
\, \frac{(1-z) \, l_\perp^2/(D-2)}{[l_\perp^2- z(1-z) \, (n_+p') (- n_-p')]\,[l_\perp^2- z(1-z) \, (n_+p') \, (\omega- n_-p')]} \,,
\cr &
\end{align}
In this case, the integral over transverse momenta is UV divergent and needs to be regularized, as indicated.
However, the divergence only influences the real part, while the imaginary part gives a similar result as
before, leading to 
\begin{align}
 \hat B \hat \Pi_\Lambda^\mu(\hat \omega_M)|_{\rm subtr.} &= - \frac{1}{8} \,
\frac{\alpha_s C_F}{2\pi} \, \hat f_{\Lambda_{b}}^{(2)}\, \gamma^\mu_\perp \, 
\tilde \Gamma  \, u_{\Lambda_{b}}(v,s) \, \int_0^{\hat\omega_s} \!\frac{d\omega'}{\hat\omega_M} \, e^{-\omega'/\hat\omega_M} 
\int_0^\infty \! d\omega  \cr 
 & \qquad \times \phi_{\Lambda_b}(\omega) \left( \theta(\omega'-\omega) + \frac{\omega'}{\omega} \, \theta(\omega-\omega') \right) 
 \,.
\end{align}
In the formal limit $\hat\omega_{s,M}\ll \omega_0$, this factorizes again, according to
\begin{align}
 \hat B \hat \Pi_\Lambda^\mu(\hat\omega_M)|_{\rm subtr.} & \simeq - \frac{1}{8} \,
\frac{\alpha_s C_F}{2\pi} \, \hat f_{\Lambda_{b}} \, \gamma^\mu_\perp \, \tilde \Gamma  \, u_{\Lambda_{b}}(v,s) \, 
\int_0^\infty \! \frac{d\omega}{\omega} \, \phi_{\Lambda_b}(\omega) 
\left( \hat\omega_M - e^{-\hat\omega_s/\hat\omega_M} (\hat\omega_M+\hat\omega_s) \right) 
 \,,
\end{align}
showing the same dependence on the sum-rule parameters as before in (\ref{eq:delxilimit}).
For the hadronic side of the sum rule, in the simplified set-up, we now find
\begin{align}
 \hat \Pi_\Lambda^\mu \big|_{\rm reson.} 
&=  \frac{\hat f_\Lambda \, M_{\Lambda_b} \, \Delta\xi_\Lambda}{m_\Lambda^2-(n_+p')(n_-p')} \, 
    \frac{\slash n_+\slash n_-}{4} \, (\slash p'+m_\Lambda) \, \gamma^\mu_\perp \, \tilde \Gamma  \, u_{\Lambda_b}(v,s) 
\cr 
&\simeq \frac{1}{n_+p'} \, 
  \frac{\hat f_\Lambda \, m_\Lambda \, M_{\Lambda_b} \, \Delta\xi_\Lambda}{m_\Lambda^2/(n_+p')- (n_-p')} \, \gamma^\mu_\perp \, \tilde \Gamma  \, u_{\Lambda_b}(v,s)  \,,
\end{align}
leading to the sum rule
\begin{align}
&  e^{-m_\Lambda^2/(n_+p')\hat\omega_M} \, \frac{\hat f_\Lambda \, M_{\Lambda_b} \, m_\Lambda}{\hat\omega_M \, n_+p'} \,  \Delta \xi_\Lambda
\cr  
& = - \frac{1}{8} \,
\frac{\alpha_s C_F}{2\pi} \, \hat f_{\Lambda_{b}}\,\int_0^{\hat\omega_s} \!\frac{d\omega'}{\hat\omega_M} \, e^{-\omega'/\hat\omega_M} 
\int_0^\infty \! d\omega  \, \phi_{\Lambda_b}(\omega) \left( \theta(\omega'-\omega) + \frac{\omega'}{\omega} \, \theta(\omega-\omega') \right) 
 \,.
\end{align}
In the simplified picture, the hard-collinear correction term $\Delta \xi_\Lambda$ could also be obtained from
the QCD factorization approach, in complete analogy to the mesonic case discussed in \cite{Beneke:2000wa}.
This will lead to an (endpoint-converging) convolution of a hard-scattering kernel
and the above LCDAs for light and heavy baryons in the di-quark approximation.
In the heavy-mass limit, the above sum-rule expression can then be interpreted as a particular model for the light-cone
wave function of the $\Lambda$ baryon, in a similar way as it has been discussed for the mesonic form factors
in \cite{DeFazio:2005dx}.


\clearpage

\end{appendices}

\renewcommand{\thefigure}{\arabic{figure}}
\renewcommand{\theequation}{\arabic{equation}}

\setcounter{figure}{0}
\setcounter{equation}{0}

\end{document}